\RequirePackage{fix-cm}
\documentclass[smallcondensed]{svjour3}     
\smartqed  

\usepackage[utf8]{inputenc}
\usepackage[T1]{fontenc}
\usepackage{graphicx}
\usepackage{amssymb}
\usepackage{natbib}
\usepackage{xcolor,soul}
\usepackage{soulutf8}

\soulregister\citep7
\soulregister\cite7
\soulregister\ref7

\PassOptionsToPackage{hyphens}{url}\usepackage{hyperref}

\begin{document}

\title{Pore-Scale Transport and Two-Phase Fluid Structures in Fibrous Porous Layers: Application to Fuel Cells and Beyond\thanks{Preprint submitted to Transport in Porous Media}
}

\titlerunning{Pore-Scale Transport and Two-Phase Fluid Structures in Fibrous Porous Layers}        

\author{Meisam Farzaneh
\and Henrik Ström
\and Filippo Zanini
\and Simone Carmignato
\and Srdjan Sasic
\and Dario Maggiolo
}


\institute{
M. Farzaneh, H. Ström, S. Sasic, D. Maggiolo\at
Department of Mechanics and Maritime Sciences, \\
Chalmers University of Technology,\\
Göteborg, SE-412 96, Sweden \and
F. Zanini, S. Carmignato \at
Department of Management and Engineering, \\
University of Padova, \\
Stradella San Nicola 3, 36100 Vicenza, Italy
\and
D. Maggiolo \at \email{maggiolo@chalmers.se}           
}

\date{Received: date / Accepted: date}

\maketitle

\begin{abstract}
We present pore-scale simulations of two-phase flows in a reconstructed fibrous porous layer. The  three dimensional microstructure of the material, a fuel cell gas diffusion layer, is acquired via X-ray computed tomography and used as input for lattice Boltzmann simulations. We perform a quantitative analysis of the multiphase pore-scale dynamics and we identify the dominant fluid structures governing mass transport. The results show the existence of three different regimes of transport: a fast inertial dynamics at short times, characterised by a compact uniform front, a viscous-capillary regime at intermediate times, where liquid is transported along a gradually increasing number of preferential flow paths of the size of one-two pores, and a third regime at longer times, where liquid, after having reached the outlet, is exclusively flowing along such flow paths and the two-phase fluid structures are stabilised. We observe that the fibrous layer presents significant variations in its microscopic morphology, which have an important effect on the pore invasion dynamics, and counteract the stabilising viscous force. Liquid transport is indeed affected by the presence of microstructure-induced capillary pressures acting adversely to the flow, leading to capillary fingering transport mechanism  and unstable front displacement, even in the absence of hydrophobic treatments of the porous material. We propose a macroscopic model based on an effective contact angle that mimics the effects of the such a dynamic capillary pressure. Finally, we underline the significance of the results for the optimal design of face masks in an effort to mitigate the current COVID-19 pandemic.
\end{abstract}

\section{Introduction}

Hydrogen is a clean and flexible energy carrier that plays an important role in the decarbonisation of the industry and transport sectors~\citep{cells2020}. As an example, hydrogen can be used in fuel cells to generate electricity, power or heat. Owing to their many advantages such as high power density, near-zero emissions and low noise, hydrogen-based Polymer Electrolyte Membrane fuel cells (PEMFC) are a reliable alternative to fossil-based fuel engines for transport, industrial and household applications~\citep{cano}.

The core of a PEMFC consists of the polymer electrolyte membrane sandwiched between two thin catalyst layers and two porous electrodes, which are called Gas Diffusion Layers (GDLs). The catalyst layer has a great impact on the cost of fuel cells and research efforts have been devoted to its improvement~\citep{brown2019}. On the cathode catalyst layer surfaces, protons and oxygen participate in an exothermic chemical reaction the products of which are water and heat. Within the pores and interstices of the GDL porous microstructure, the reactant gas and the produced water flow to and from the catalyst layer. The GDL is the key component for regulating the transport phenomena inside the cell and, in turn, its design is an important factor to determine cell performance~\citep{majlan,balakrishnan}. In particular, the GDL porous microstructure is fundamental in determining the two-phase (liquid-gas) dynamics and distribution in the cell, a complex mechanism of mass transport that affects the management of the produced water~\citep{berg}.

Managing the water balance inside fuel cells is a great challenge. An optimal operation of a fuel cell requires an adequate hydration of the membrane for allowing proton conductivity. On the other side, excessive water fills up the pores and interstices of the GDL, blocking the access of reactant gas to the surfaces of the catalyst layer and decreasing the available active sites for reaction~\citep{culubret}. Consequently, the accumulation of produced water, which is mostly observed at high current densities, limits cell performance~\citep{huili}. The imbalance of water inside the cell is particularly important in new types of fuel cells, such as anion exchange membrane fuel cells~\citep{eriksson,chen2020}. An inefficient water management in fuel cells is also recognized as one of the leading mechanisms that induce fuel starvation and cell degradation -- two well-known ongoing issues that reduce the lifetime of fuel-cell systems~\citep{chen2019,nandjou2016}. It is therefore of great interest to investigate the interplay between two-phase flow phenomena and GDL microstructure, in order to establish efficient water management strategies, and increase the performance, robustness and competitiveness of fuel cells in the energy market.

GDLs are composed of a highly porous (porosity $>70 \% $) woven carbon cloth or carbon paper with $100-500\ \mu m$ thickness. The diameter of fibres composing the GDLs is typically $10-15\ \mu m$ and the mean pore size is $50-120\ \mu m$~\citep{gostick2007,lu2010}. The GDL porous microstructure provides a complex network of narrow and dilated pore throats giving rise to local constrictions of the flow paths. This complex topology induces high curvatures of the liquid-gas interface and, given the presence of surface tension, it produces significant capillary forces that act as a resistance to the flow (see e.g. \citet{lu2019}).
The presence of these capillary forces is a source of pressure buildup and intermittent water dynamics in the GDL. This behaviour often results in impulsive forces that flush out the water to restore the cell operational performance~\citep{huili}.

The strongly intermittent transport phenomena in GDL samples have been experimentally observed. As an example, fluctuations of the pressure drop, induced by the breakdown and re-build of water flow paths, have been measured~\citep{mortazavi2014,niknam}. Water can emerge from different preferential locations on the porous surface when transported through a GDL, an observation that suggests that the spatial distribution of flow paths is also intrinsically dynamic~\citep{lu2010}. The observed behaviours of liquids penetrating into GDLs are consistent with the physical description of two-phase flows in porous media, as characterised by intermittent fluid connectivities that are strongly related to the porous microstructure~\citep{reynolds}. Recent advances in imaging analysis techniques have allowed the access to pore-scale flow visualization within fuel cells, for instance by means of Nuclear Magnetic Resonance (NMR) imaging, neutron imaging and X-ray computed tomography (CT). Even though these experimental methods provide valuable information on microscale water transport behaviour within fuel cells, their application is often challenging due to opacity of the system, limited spatial and temporal resolutions, cost and availability~\citep{bazylak2009}.

Numerical simulations provide a valuable tool for overcoming the limitations related to experimental techniques, in terms of both investment and temporal costs~\citep{andersson2016}. As an example, the two-phase flow in fuel cells has been mathematically modelled and simulated using continuum approaches based on the volume averaging of physical quantities~\citep{nam2003}. However, GDLs suffer from the lack of length scale separation since the thickness of a GDL measures only few pore sizes, making the application of continuum models controversial~\citep{rebai2009}. To incorporate the GDL pore-scale structure in numerical simulations, Pore Network (PN) models~\citep{lenormand1988, rebai2009} and artificially generated fibrous material have been used~\citep{chen2012,deng2019,hao2010}.
PN models represent the microstructure of porous media as a  two- or three-dimensional network of idealized pores and throats, whereas artificial materials are usually created by means of stochastic techniques  based on macroscopic design specifications, such as porosity and fiber size.
Such studies can catch the small-scale transport mechanisms usually filtered out in macroscopic continuum models but they rely on simplified microstructures. Therefore,  pore-scale simulation of liquid water transport and distribution inside realistic GDLs geometries can unveil a much richer physical picture.
An interesting approach for the investigation of pore-scale two-phase dynamics makes use of X-ray computed tomography for reconstructing porous media as input for microscale simulations (see e.g. ~\cite{raeini2014}). Some studies have exploited this strategy  focusing on reconstructed GDL geometries ~\citep{rama2012,gao2013}. These studies investigated the three-dimensional  evolution of liquid invasion in hydrophilic and hydrophobic GDLs under different pressure gradients.
The prominent effect of microstructure on two-phase fluid distribution and dynamics has been observed in samples with surface areas of the size of only few pores, i.e. of the order of a hundred micrometers squared. These insights depict the significance of such analyses and motivate further investigations of two-phase flows in GDLs, applying the analysis to samples of larger size. Indeed, a comprehensive analysis of the pore-scale two-phase transport in GDL samples is still lacking, thus limiting the possibility of understanding the microscopic behaviours of the fluids inside the fuel cell.

Recent studies have shown the potential of combining lattice Boltzmann simulations and X-ray computed tomography for the investigation of single-phase flows in large samples of porous battery electrodes, notably providing the computation of electrode conversion and reactive properties~\citep{kok,maggiolo2020}.
In the present study, we adopt a similar approach to perform pore-scale two-phase lattice Boltzmann simulations in the porous structure of a commercially available carbon paper GDL for fuel cells (AvCarb MGL370), reconstructed via X-ray computed tomography. Such fibrous layer materials are characterised by much larger lateral dimensions compared to the thickness. The pore-scale heterogeneities are expected to be statistically significant along the minor dimension, which is of the size of a few pores, and to play an important role in the global performance of the material. We are particularly interested in the effect that the microstructural variations have in determining the dynamics along the medium thickness.

Three samples with a surface area and thickness of the order of 1 square millimetre and 200 $\mu$m, respectively, are extracted from the reconstructed materials and used as input for three-dimensional two-phase flow simulations to mimic a liquid-gas (water-vapour) system.
The results of the simulations  are presented in terms of liquid water dynamics, spatial distribution and flow preferential paths. We capture the two-phase flow dynamics within the GDL at the pore-scale and we illustrate that the flow experiences a complex dynamics in the form of different spatial patterns, where inertial, capillary and viscous forces all contribute to the intermittent flow behaviour. These results provide important information for expanding the knowledge about the physical mechanisms underlying the intermittent mass transport phenomena in fuel cells, which often cause fluctuation in the performance and unstable operation of a fuel cell~\citep{ji2009,stpierre2000}.
Finally, we also propose a model to describe the water transport in the GDL based on extracted effective permeability values.

The results on the two-phase flow dynamics and fluid structures, and their relation to the porous microstructure, provide new insight on the pore-scale mechanisms of transport in fibrous materials, also beyond the application to fuel cells. In the final part of the paper, we will share important aspects regarding the optimal design of face masks with an improved filtering performance, and we discuss how the use of certain type of materials in the production of personal protective equipment can possibly contribute to limit the transmission of viruses and mitigate the current COVID-19 and future pandemics.

The manuscript is organized as follows: in Section~\ref{sec:numet} the material reconstruction technique and the numerical methodology for simulating two-phase flows are presented; in Section~\ref{sec:res} we present and discuss results of the numerical analysis; in Section~\ref{sec:covid} we discuss the relevance of the results for the design of face masks devoted to the mitigation of virus transmission, and in Section~\ref{sec:end} we summarize the main findings of the research.

\section{Numerical Methodology\label{sec:numet}}

\subsection{GDL reconstruction via X-ray computed tomography\label{sec:mat}}

\begin{figure}
    \centering
    \includegraphics[width=7cm]{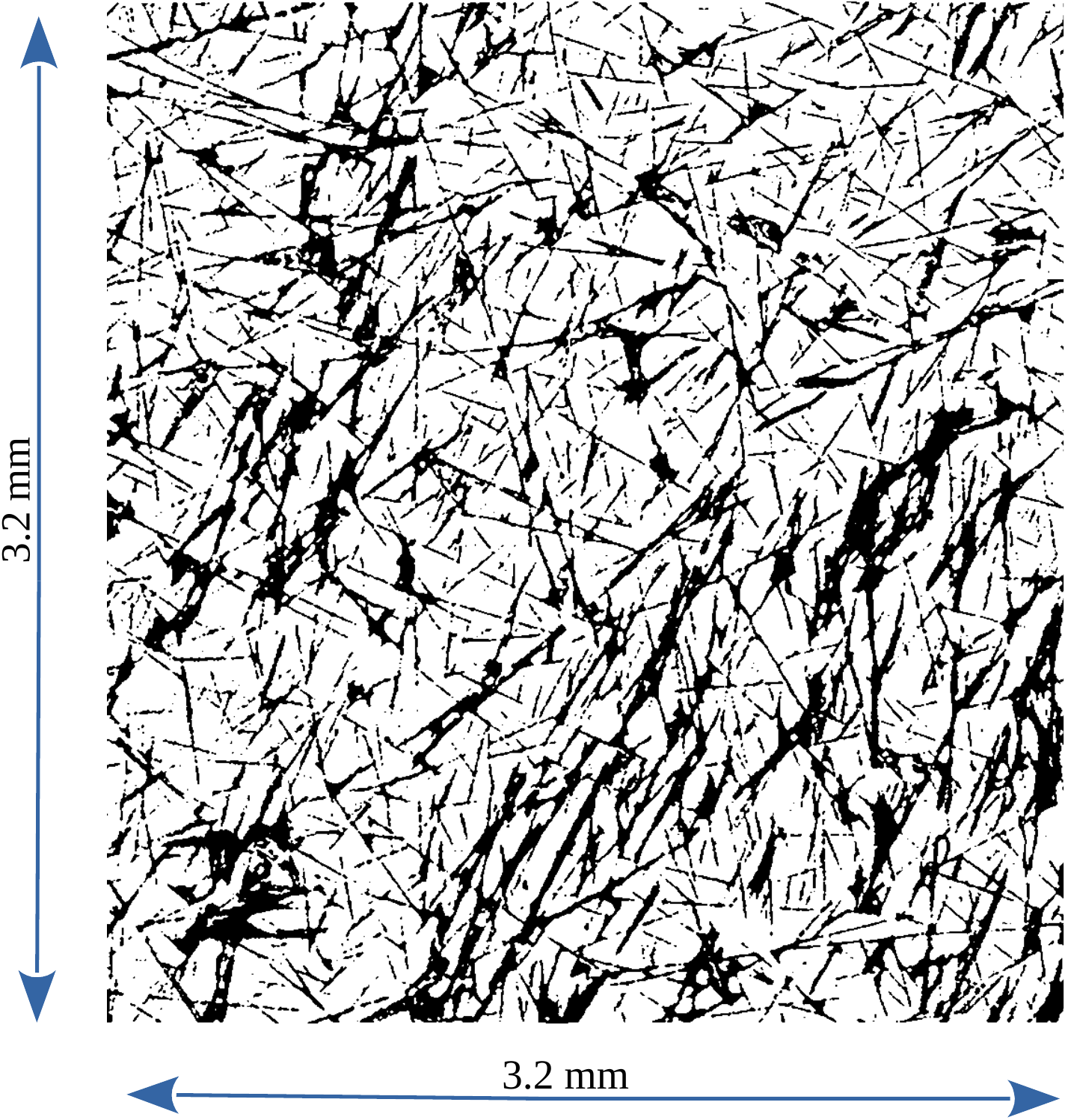}
    \caption{A cross-sectional view extracted from the CT reconstructed volume  and binarised to discriminate between material (black) and pores (white). The size of the total surface area is $\sim 3.2 \ mm \times 3.2 \ mm$}
    \label{fig:scan}
\end{figure}

The three-dimensional geometry of the GDL material (AvCarb MGL370) is acquired and reconstructed via X-ray CT. For that purpose, a metrological CT system (Nikon Metrology MCT225)  is used, which is characterised by
micro-focus X-ray source (minimum focal spot size equal to 3 $\mu m$), 16 bit detector with 2000x2000 pixels and cabinet ensuring controlled temperature of 20 $^\circ$C. The used CT scanning parameters are reported in Table~\ref{tab:ct}.
In order to effectively acquire the geometry of micro-scale fibres, the CT metrological structures resolution has to be maximised~\citep{zanini2017two}. To this end, the voxel size is minimised by reducing the relative distance between the sample and the X-ray source. This is possible because small portions are cut out from the initial material, with dimensions of about $3.2 \ mm \ \times 3.2\ mm \ \times 0.37\ mm$. Since the thickness (0.37 $mm$) is lower than the other dimensions, eight samples are packed and scanned with a single acquisition procedure. As a consequence, the scanning time reported in Table~\ref{tab:ct} is the time needed to scan up to 8 material portions.
In addition, to further improve the scanning resolution, the focal spot size is also minimised by keeping the X-ray power below 7 Watts. A large number of bi-dimensional projections (i.e. gray-scale images representing the local attenuation of X-rays occurring when traversing the samples material) are acquired at a different angular position of the samples during their revolution around the rotary table axis. Starting from these projections, a three-dimensional reconstruction is conducted using a filtered back-projection algorithm~\citep{feldkamp1984} and cross-sectional images (as the one seen in Fig.~\ref{fig:scan}) are extracted for each sub-sample (with the distance between consecutive sections equal to the voxel size), exported into image stacks and binarised to discriminate between material (black) and pores (white).

\begin{figure}
    \centering
    \includegraphics[width=0.89\linewidth]{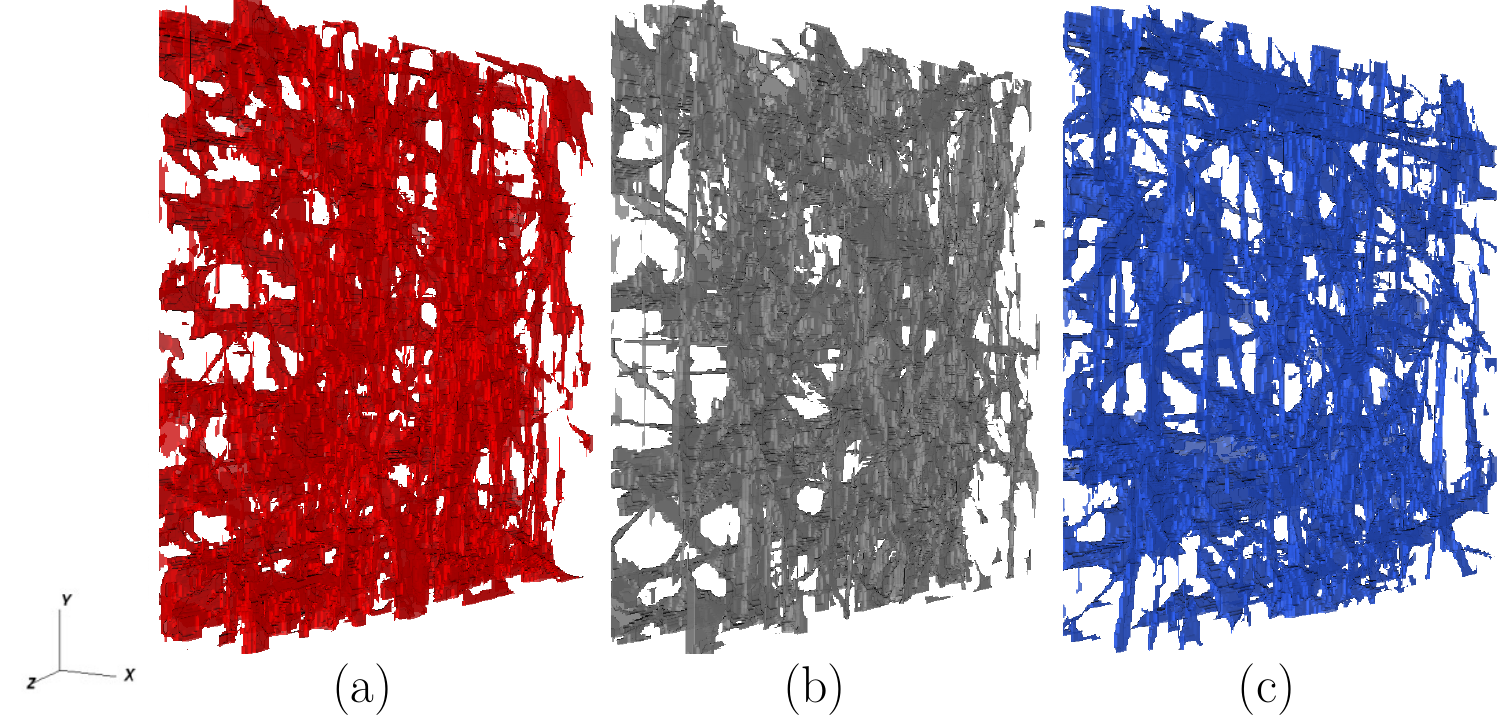}
    \caption{Illustration of the 3 portions of the CT reconstructed GDL geometry (REVs) selected for flow simulations. The size of the REVs is characterized by a surface area of $A=819.2 \ \mu m \times 819.2 \ \mu m$ and thickness $L_{pm}=192\ \mu m$}
    \label{fig:scan3d}
\end{figure}

\begin{table}
\caption{The CT scanning parameters.~\label{tab:ct}}
\begin{tabular}{c  c  c }
\hline\noalign{\smallskip}
Parameter & Value & Unit \\
\noalign{\smallskip}\hline\noalign{\smallskip}
Voltage & 80 & $kV$ \\
Current  & 83 & $\mu A$ \\
Power & 6.6  & $W$ \\
Exposure time & 4000 & $ms$ \\
Nr. of projections & 2000 & -- \\
Voxel size  & 3.2  & $\mu m$ \\
Scanning time & 133 &  $min$ \\
\noalign{\smallskip}\hline
\end{tabular}
\end{table}

Even if the resolution is maximized, the smallest fibres may present discontinuous shapes (as seen in Fig.~\ref{fig:scan}), because their microstructure
cannot be fully captured by the CT scan. For this reason, a filtering procedure is operated, based on the cubic interpolation of three-dimensional geometry in a grid with increased resolution. Following this image analysis technique we obtain an enhanced three-dimensional representation of the material, with a final resolution of $1.6 \ \mu m$. Compared to the original material's specifications, we obtain a slightly higher porosity, from 78\% to 80\%. We also have to reduce the thickness of the sample, from the original  $370\ \mu m$ to $192\ \mu m$ in order to avoid unphysical effects at the borders due to the presence of adjacent materials at the two sides of the GDL during the CT acquisition.

\begin{figure}
    \centering
    \includegraphics[width=8cm]{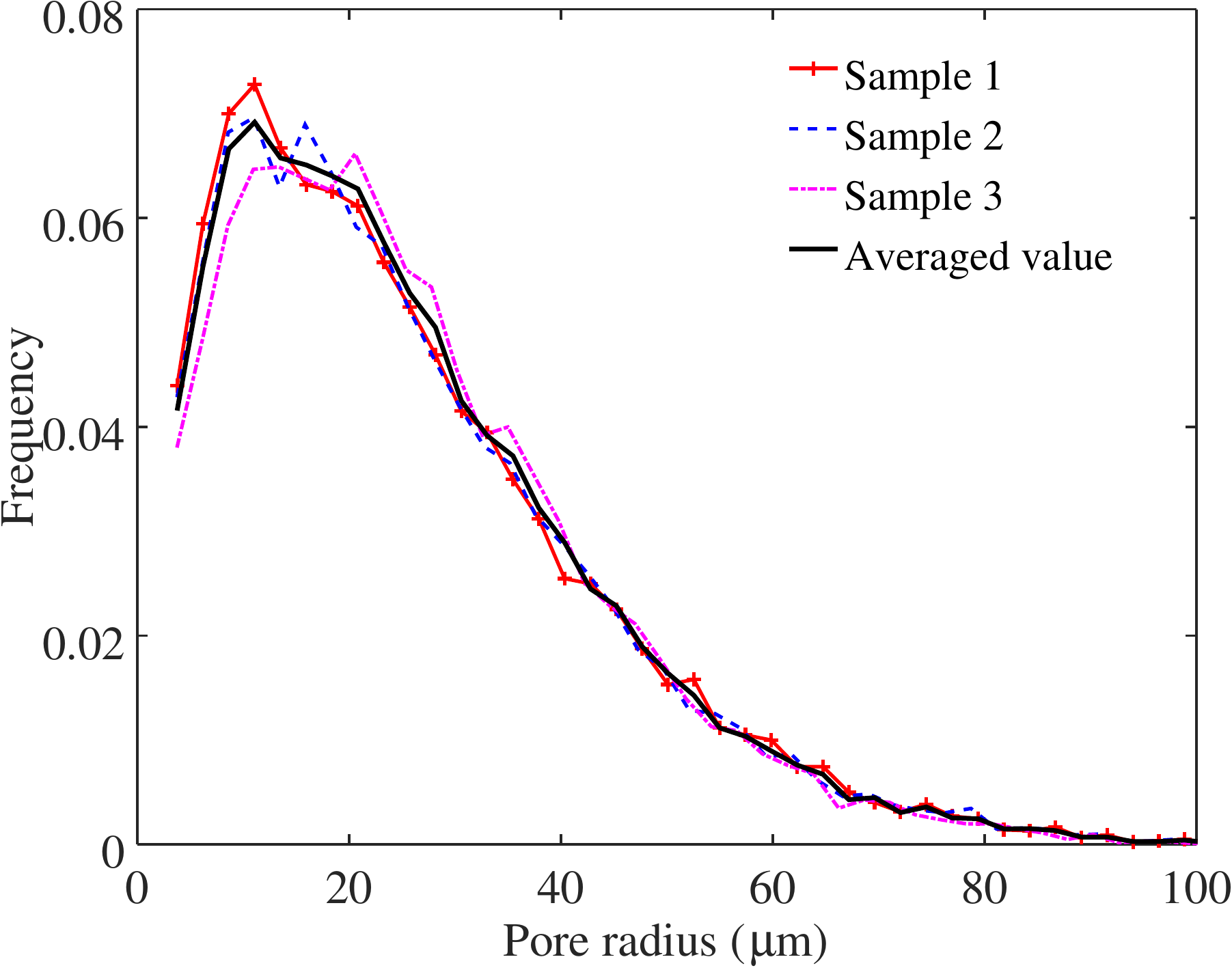}
    \caption{Probability distribution function of pore diameters for the three REVs used as input for numerical simulations and averaged values. The pore size distributions are calculated using the watershed algorithm presented in~\citep{Rabbani2014}. It is observed that all the three REVs have similar pore size characteristics}
    \label{fig:psd}
\end{figure}

Subsequently, we select three portions (representative element volumes, REVs, Fig.~\ref{fig:scan3d}) of the reconstructed GDL to serve as input for the numerical simulations. The three REVs are chosen in order to increase the statistical confidence of the simulation data. The computational domain of each sample is then characterized by a cross-sectional surface area of $819.2 \ \mu m \times 819.2 \ \mu m$ and a thickness of $192\ \mu m$, to which correspond $512 \times 512$  and $120$ computational cells, respectively.
The pore size distributions of each materials samples are presented in Fig.~\ref{fig:psd} where we also depict the pore size distribution averaged over the 3 samples. We observe that the material has a microscopical structure characterized by pore sizes that range from a few microns to tens of microns. The average pore radius results $r_p=40\ \mu m$, on the basis of which we argue that
the size of the domain, with a surface area approximately $10^2$ times larger than the average pore size, well captures the
spatial variation that characterizes the GDL microstructure.

\subsection{Lattice Boltzmann Method for two-phase flow simulations}

The lattice Boltzmann method is a numerical approach that allows the study of multiphase flows in porous media at a low computational cost~\citep{zhao2019}. It is well suited for numerical solutions of complex flows with pore-scale accuracy and it is therefore able to capture the two-phase fluid-dynamic behaviors determined by the material microstructural characteristics of a fuel cell GDL. The GDL geometry, reconstructed via CT  and filtered as explained in Section~\ref{sec:mat}, is here used as input for pore-scale simulations.
We make use of a single-component two-phase flow model to mimic a liquid-gas (water-vapour) system by means of the numerical methodology proposed by Shan and Chen~\citep{shanchen} for the simulation of surface tension effects.  Such a methodology does not rely on the dynamic solution of the two-phase interface. Instead, the surface tension force acting between the two phases is simulated by Van-der-Waals-like intermolecular interactions.

The two-phase flow is solved on a three-dimensional regular Cartesian grid (D3Q19) by means of the local solution of the probability distribution function $f_r$, a statistical quantity that indicates the molecular probability of fluid mass and momentum~\citep{succi}. The lattice Boltzmann equation reads as:
\begin{eqnarray}
f_{r}(\mathbf{x}+\mathbf{c}_{r}\delta t, t+\delta t)-f_{r}(\mathbf{x},t) =  
-\tau^{-1}(f_{r}(\mathbf{x},t)-f_{r}^{eq}(\mathbf{x},t))+F_{r}
\label{eq:eq1}
\end{eqnarray}
where $\mathbf{x}=(x,y,z)$ is the position vector, $t$ is the time, $\mathbf{c}_{r}$ is the so-called discrete speed, $f_{r}^{eq}$ is the equilibrium distribution function, and the subscript $r=1...19$ indicates the lattice directions. The relaxation time $\tau$ is directly linked to the fluid dynamic viscosity as $\mu(\mathbf{x},t)=\rho c_s^2 (\tau-0.5)$, where $c_s$ is the speed of sound and $\rho=\rho((\mathbf{x},t)$ the fluid density. The fluid is forced by an applied body force $F_{r}$ that mimics the effect of a pressure gradient $-\nabla_x P$ acting on the $x$ direction.  The body force is defined as \citep{Guo2002}:
\begin{equation}
\label{eq:eq2}
F_{r}=\left( 1-\frac{1}{2\tau} \right)w_{r}\left( \frac{\mathbf{c}_{r}-\mathbf{u}}{c_{s}^2}+\frac{\mathbf{c}_{r}\ \mathbf{u}}{c_{s}^4} \mathbf{c}_{r} \right) \ (-\nabla_x P)
\end{equation}
where $w_{r}$ represents the weighting coefficient and $\mathbf{u}=\mathbf{u}(\mathbf{x},t)$ the Eulerian velocity vector. It should be noted that in our simulations the pressure gradient $\nabla_x P$ acts only in the streamwise direction $x$, whereas it is null along the transverse directions $y$ and $z$ (see also the sketch in Fig.~\ref{fig:geo}). The value of the applied pressure gradient is discussed in the next subsection.

The equilibrium distribution function is defined along the $r$-th direction as:
\begin{eqnarray}
f_{r}^{eq}(\mathbf{x},t) &=&w_{r} \rho \left( 1-\frac{\mathbf{u}_{eq}  \mathbf{u}_{eq}}{2c_{s}^{2}} \right), \ r=1 \\
f_{r}^{eq}(\mathbf{x},t) &=& w_{r} \rho \left( 1 + \frac{\mathbf{c}_{r} \mathbf{u}_{eq}}{c_{s}^{2}}    +  \frac{(\mathbf{c}_{r}  \mathbf{u}_{eq})^{2}}{2c_{s}^{4}}  \right ),   \ r=2-19 \ ,
\label{eq:eq7}
\end{eqnarray}
where $\rho \mathbf{u}_{eq} = \rho \mathbf{u} + (\tau - 1/2) F_{sc}$~\citep{chen2014}. The interaction force between different phases $F_{sc}(\mathbf{x},t)$ is based on a density-dependent pseudopotential function $\Psi(\rho)$ \citep{shanchen}:
\begin{equation}
\label{eq:eq3}
F_{sc}(\mathbf{x},t)=-G\Psi(\mathbf{x},t) \sum\limits_{r} w_{r} \Psi(\mathbf{x}+\delta x,t)\mathbf{c}_{r}
\end{equation}
where $G$ determine the strength of interaction between the two phases ($G=-5.5$ in the present model). The form of the psudopotential function, defined as $\Psi(\rho)=1-e^{-\rho}$, yields a non-ideal equation of state that allows the coexistence of the two phases:
\begin{equation}
\label{eq:eq4}
P(\rho)=\rho c_{s}^{2}+\frac{G}{2}c_{s}^{2}\Psi(\rho)^{2} \ .
\end{equation}
One of the known limitations of the two-phase lattice Boltzmann models is the emergence of numerical instabilities and spurious currents in cases with high density and dynamic viscosity ratios between the phases. In the present study, the dynamic viscosity ratio is set as $M=\mu/\mu_g=35$ (with $\mu$ and $\mu_g$ the viscosities of liquid and gas, respectively), a value that we found  as a good compromise between accuracy of the solution ($M=100$ for water-air systems) and numerical stability. The reader is referred to the work presented in \citet{pettersson2020} for the two-phase flow algorithm validation.

The fluid-wall interaction at the three-phase contact line (liquid-gas-solid), in terms of chemical properties of the surface (wetting), can be tuned by means of an averaging of the pseudopotential values in the neighboring nodes $N$:
\begin{equation}
\label{eq:eq5}
\Psi_{wall}(\rho)=\Psi \ \Big ( N^{-1}\sum\limits_{N} \rho+\Delta_{w} \Big )
\end{equation}
The resulting contact angle is determined by the value of the parameter $\Delta_{w}$~\citep{demaio2011,vaitukaitis2020}. The contact angles in fuel cells GDLs have been measured to widely vary, over a range $\theta \approx 80-150^\circ$ on the basis of temperature and hydrophobic agent content~\citep{lim2004}. The value of the contact angle is known to affect fuel cells performance (see e.g.~\citet{park2009}) but less is known about the effect of the porous microstructure. Both wettability and medium geometrical disorder are expected to greatly impact fluid displacement in porous media (see e.g. \citet{wang2019}), thus motivating the need of a deeper understanding of the effects of pore-scale microstructural traits on multiphase flows in fuel cells. In the present study, we address such a scientific question by simulating the two-phase flow dynamics within a GDL without hydrophobic treatments, and the equilibrium contact angle is set to $\theta_{eq}=90^\circ$.

Finally, the macroscopic fluid density and momentum are determined through the statistical averaging of the molecular probability distribution functions as~\citep{succi}:
\begin{eqnarray}
\rho &=&\sum\limits_{r} f_{r}(\mathbf{x},t)\\
\rho\mathbf{u} &=& \sum\limits_{r} f_{r}(\mathbf{x},t)\mathbf{c}_{r}+\frac{1}{2}(-\nabla_x P) + \frac{1}{2} F_{sc} \ .
\label{eq:eq6}
\end{eqnarray}

\subsection{Simulations setup\label{sub:setup}}

\begin{figure}
    \centering
    \includegraphics[width=0.6\linewidth]{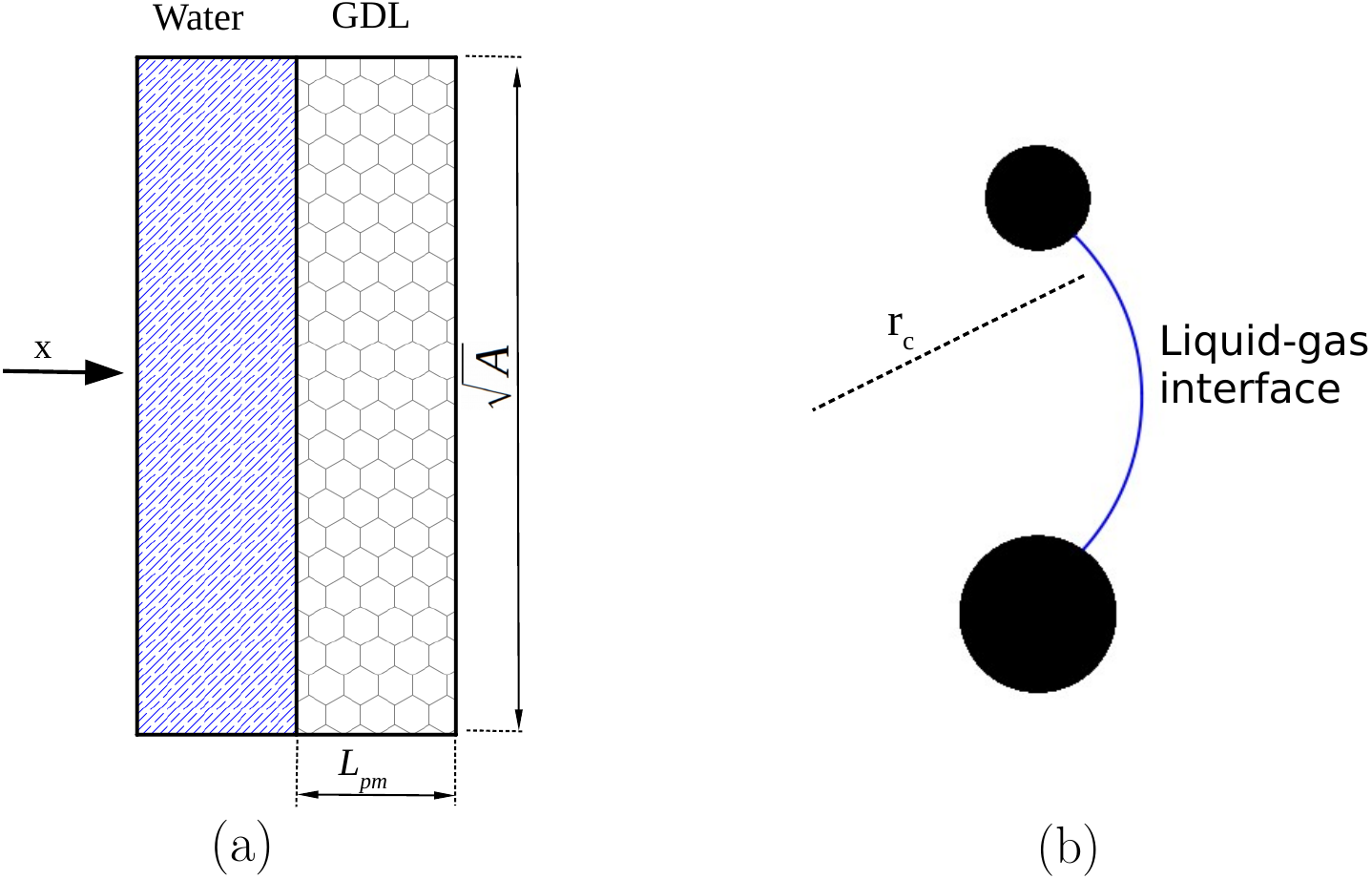}
    \caption{(a) Schematic of the computational geometry in the present work; The flow cross section $A$ is $819.2 \mu m \times 819.2 \mu m$ and the GDL thickness $L_{pm}$ is $192 \mu m$ (b) Representation of the liquid front penetrating through the space between two fibers and the induced liquid-gas interface curvature of radius $r_c$. The value of $r_c$ can be approximated with the pore throat radius, i.e. $r_c\approx r_p$}
    \label{fig:geo}
\end{figure}

We investigate the dynamic infiltration into the GDL of a volume of water $V\sim 0.13 \ mm^3$.
Such a volume is equal to the total volume of the investigated porous samples $AL_{pm}$, so that we ensure that all the media can possibly be filled with water, i.e. a situation where the cell is experiencing high flooding. A sketch of the computational domain is presented in Fig.~\ref{fig:geo}(a). Along the streamwise $x$ direction we impose a pressure gradient that drives the water through the GDL microstructure and periodic boundary conditions at the domain boundaries. At the boundaries along the transverse directions $y$ and $z$, the symmetry is
guaranteed by imposing free-slip boundary conditions.

When the water produced on the catalyst layer approaches a pore throat, such as the constriction induced by intersecting fibers (see e.g. Fig.~\ref{fig:geo}(b)), it is temporary held up and its transport impeded until the local water pressure exceeds the capillary pressure induced by the liquid-gas interface curvature. The competition between the values of the pressure experienced by the water and the pressure threshold induced by the pore geometry is the mechanism that determines if the pore throat is infiltrated or not. Based on the mean pore radius of our sample, the capillary pressure threshold can be in the first instance estimated as $P_t\sim 2 \sigma/r_{p} = 3.6 \ kPa$ (with $\sigma = 0.072 \ N/m$ the water-air surface tension). Such a value is indeed consistent with experimental observations of water penetrating a GDL and with the corresponding measured pressure buildup~\citep{mortazavi2014}.

We then assume that the driving force for water motion is the pressure gradient $-\nabla_x P$ determined by the difference between the pressure buildup at the GDL entrance (at the catalyst layer-GDL interface) and the pressure of the gas phase at the end of the porous domain (at the GDL-gas channels interface). The gas channel is assumed to be  at atmospheric pressure. Following this reasoning, the pressure gradient results $-\nabla_x P = P_t/L_{pm} = -187 \ kPa/cm$. This is the value of the pressure gradient that forces the flow in our simulations, implemented through equations~\ref{eq:eq1} and~\ref{eq:eq2}. We neglect gravitational effects, since in fuel cells applications the associated volumetric force $\rho g\sim0.1 \ kPa/cm $ results three orders of magnitude lower.
It should be noted that the value of the water pressure at the catalyst layer-GDL interface can vary greatly during cell operation. For instance, such a variation can be due to the intermittent production of water that a cell experiences or to the complex effects that microstructural and chemical properties of the catalyst layer induce on the thermodynamical properties of water~\citep{eikerling1998}. The presence of a microporous layer in between the catalyst layer and the GDL can complicate the picture even more.
However, the water production and behavior in the catalyst layer are out of the scope of this study. Since we are interested in following the time-dependent dynamics of water induced by the pressure buildup in the cell electrodes, we consistently apply the pressure conditions relevant to such an investigation. Under such conditions, as simulations will show, liquid penetrates in some pores of the medium and will accumulate in others, producing a rich spatial and temporal behavior that unveils important insights about water transport in fuel cell GDLs.

\section{Results\label{sec:res}}

\subsection{Water infiltration and two-phase flow dynamics\label{sec:liq}}

\begin{figure}
    \centering
\includegraphics[width=0.8\linewidth]{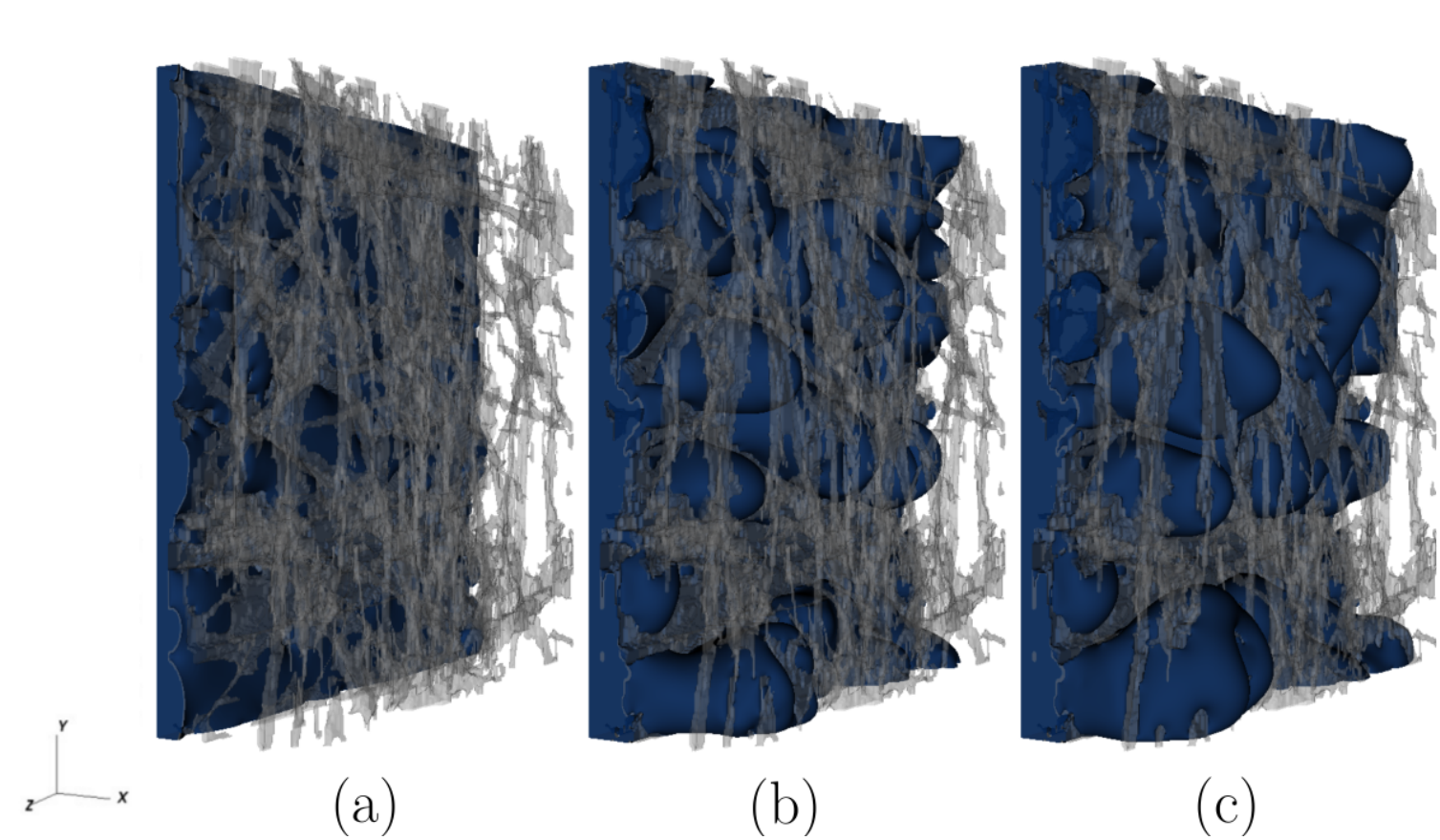}
\caption{Three snapshots of the displacing liquid water from lattice Boltzmann simulations  in sample 3:
(a) $t^{*}=0.2$, (b) $t^{*}=0.5$ and (c) $t^{*}=0.8$. The visualization illustrates that the liquid phase is squeezed into the microscopic void space of the fibrous material and transported through the porous medium microstructure. The grey coloured structure represents the solid fibers. At time instant $t^{*} \sim 0.5$, the breakthrough of the liquid-water front occurs and water emerges from the surface of the GDL} \label{fig:invasion}
\end{figure}

We first qualitatively look at the dynamics of water infiltrating the reconstructed GDL. Figure~\ref{fig:invasion} shows an example of water infiltration for one of the three investigated samples, at three different times. We observe the predicted heterogeneous behavior, with some pores filled much faster than others, and the presence of pronounced curvatures of the liquid-gas interface induced by the geometrical constraints of the porous medium. We also observe that at time $t^*\sim 0.8 $ the liquid phase has reached the GDL outlet, emerging in the form of spatially distributed fingers.

\begin{figure}
    \centering
    \includegraphics[width=0.7\linewidth]{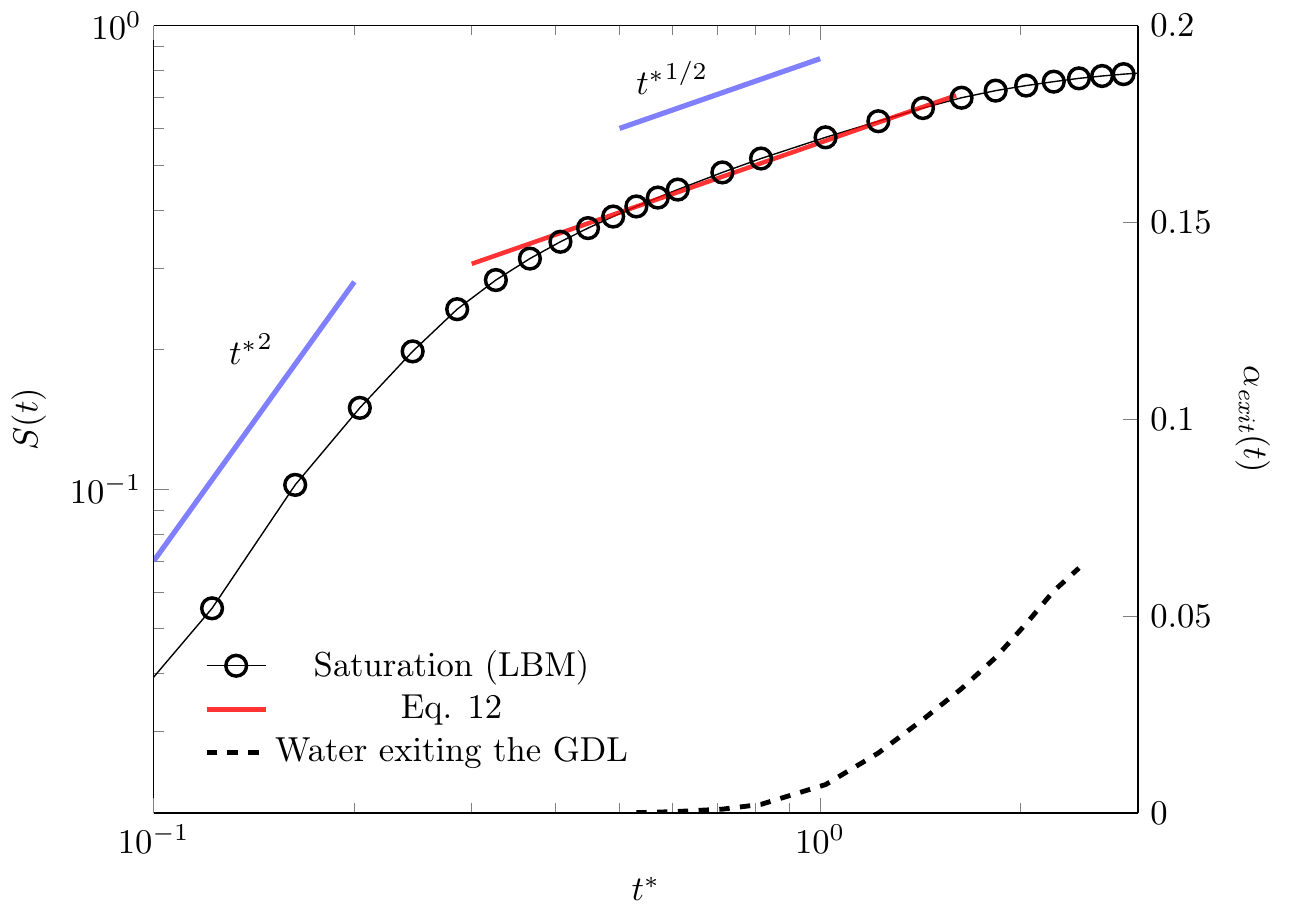}
    \caption{Water saturation as a function of time (left ordinate axis). The liquid front initially propagates with a rate $ \propto {t^*}^{2}$, due to inertial forces, whereas at the later stages, when viscous effects becomes important, the liquid penetrates $\propto {t^*}^{1/2}$. The liquid saturation at long times is fitted by Eq~\ref{eq:capillary_non}. The figure also reports on the right ordinate axis the cumulative fraction of the liquid emerging from the GDL outlet surface, $\alpha_{exit}=V_{exit}/V$ (dashed line), with $V$ the total liquid volume in the numerical domain. The latter plot is showed till $2.5{t^*}$, when the water that emerges from the outlet is reconnected with the inlet liquid front due to the periodic boundary conditions}
    \label{fig:pipe}
\end{figure}

From a macroscopic perspective, the liquid infiltration can be quantified by the percentage of volume of pores invaded by water at each instant, i.e. by the porous medium saturation $S(t)$. In Fig.~\ref{fig:pipe} we report the variation of saturation with time. For this figure, as for the other ones throughout the paper, the computed quantities refer to the average value among the three different REVs (see also Section~\ref{sec:mat}). For comparative purposes, we non-dimensionalise the time as $t^{*}=t/t_{c}$, where $t_c=L_{pm}/U_c$ is the time that the considered liquid would take to fill a tube of radius $r_p$ and length $L_{pm}$, when it experiences an average characteristic velocity $U_c=-\nabla_x P\ r_p^2 / (4 \mu_w)$ (i.e. the Poiseuille's velocity averaged in time during infiltration of a liquid in a pipe at a constant pressure drop). From this first comparison, we observe that at the latter characteristic time, i.e. when $t^*=1$, the porous medium is not completely filled and $S(t^*=1)<1$. At a time $t^*\sim 0.5$ we observe the appearance of the first liquid cluster on the GDL outlet surface, a time that we herein refer to as the breakthrough time. At a time $t^*\sim 1$ roughly 1\% of the total water has emerged along certain preferential flow paths. In the next subsection we will further discuss the mechanism that determines the spatial distribution of these preferential flow paths.

Fig.~\ref{fig:pipe} also shows that the liquid dynamics is characterised by three distinctive regimes: (i) a ballistic one where the saturation increases as $S(t)\propto {t^*}^2$, (ii) a less pronounced infiltration dynamics, where saturation increases with the square root of time, $S\propto {t^*}^{1/2}$, and (iii) a final stage where the saturation approaches a constant value. This behaviour points out that the initial infiltration is inertia-dominated, with the average liquid penetration height $h(t) =S(t)  L_{pm}$ advancing with an increasing velocity $\mathrm{d} h/ \mathrm{d} t$. We note that this inertial regime occurs when the liquid invades the initial part of the GDL (approximately the 20\%) and it can also be affected by the water dynamics in the catalyst layer. We then argue that in the presence of high pressure buildups at the catalyst layer-GDL interface, this mechanism can significantly impact the water transport in fuel cell GDLs.

When the liquid penetrates the porous medium, the amount of fibre surface wetted by the liquid increases and, in turn, the viscous resistance becomes increasingly important.
At longer times the macroscopic liquid transport is then determined by the competition between the pressure drop $-\nabla_x P\ L_{pm}$ and the viscous forces acting on the liquid $\mu_w h$. The macroscopic liquid infiltration rate does not increase, but it rather decreases proportionally to $h$:
\begin{equation}
\label{eq:darcy}
\frac{\mathrm{d} h}{\mathrm{d} t} \propto -\frac{\nabla_x P\ L_{pm}}{\mu_w h} \ .
\end{equation}
Such a competition between the pressure and viscous forces leads to the scaling $S\propto {t^*}^{1/2}$, which is the classical result of Darcy's law for a liquid infiltrating a porous medium (that is, the equivalent of the Lucas-Wahsburn solution for a flow driven by a pressure drop). We observe a direct transition from the initial inertial regime $S\propto {t^*}^2$ to the viscous regime  $S(t)\propto {t^*}^{1/2}$, and thus a lack of the often observed convective regime $S\propto {t^*}$. This kind of a transition is due to the lack of separation of the time scales characterising the inertial time $t_i=\sqrt{\rho r_p^3/\sigma}$ and the viscous time $t_v=r_p^2/\nu$, that is, to the relatively high value of the Ohnesorge number $Oh=t_i/t_v=\sqrt{\rho \nu^2 / (\sigma \ r_p)} \approx 10^{-1}$. This observation points out that the viscous forces start to play a dominant role at relatively short times and that the convective regime, which is usually observed at $t_i<t<t_v$, occurs in a short time interval~\citep{stange2003}. Since in such a short period preceding the viscous regime, an inertial-viscous transition regime is also expected~\citep{fries2008}, we deduce that the convective, purely inertial regime cannot be observed here.

After a time $t^*\ge 2$ the saturation growth rate is reduced and tends asymptotically to zero, $\mathrm{d} S/\mathrm{d}t\rightarrow 0$. As we will see in the next section, the transport behaviour at long times is in fact characterised by the occurrence of preferential flow paths that connect the inlet and outlet of the medium, and the global mechanism of medium filling is inhibited.

The inertial and viscous forces are not the only forces contributing to water transport. As we previously argued, the microstructural traits of the GDL induce strong curvatures of the liquid-gas interface and give rise to capillary forces. We measure the time-dependent average capillary pressure $P_c(t) =P_l(t) -P_g(t) -P_v(t)/2$ , as the difference between the spatially averaged liquid and gas pressures $P_{l,g}(t)=1/V_{l,g}\int_{V_{l,g}} p(\mathbf{x},t)\ \mathrm{d} V$ (with $V_l$ and $V_g$ the liquid and gas volumes, respectively), minus half of the estimated viscous pressure drop in the liquid from Darcy's law $P_v(t)=u_l\ \mu h(t)\ 4 / r_p^2$ (with $u_l$ the liquid-phase velocity). For this estimation, we assume a linear decrease of the pressure along the flow direction because of the viscous forces in the liquid, and we also assume negligible the gas viscous forces; following these assumptions we can write $P_c(t)+P_v(t)=P_{in}(t)-P_{out}(t)$, with $P_{in}(t)$ and $P_{out}(t)$ indicating the pressures at the inlet and outlet of the medium, and derive  $P_c(t)= P_{in}-P_v-P_g = P_l - P_v/2 -P_g$. A positive value of $P_c$ indicates that the liquid-gas interface is convex with respect to the streamwise direction $x$ and it represents a resistance to the flow. On the other side, a negative or a null value point to a concave or flat interface and do not provide resistance to the flow.

\begin{figure}
    \centering
    \includegraphics[width=0.9\linewidth]{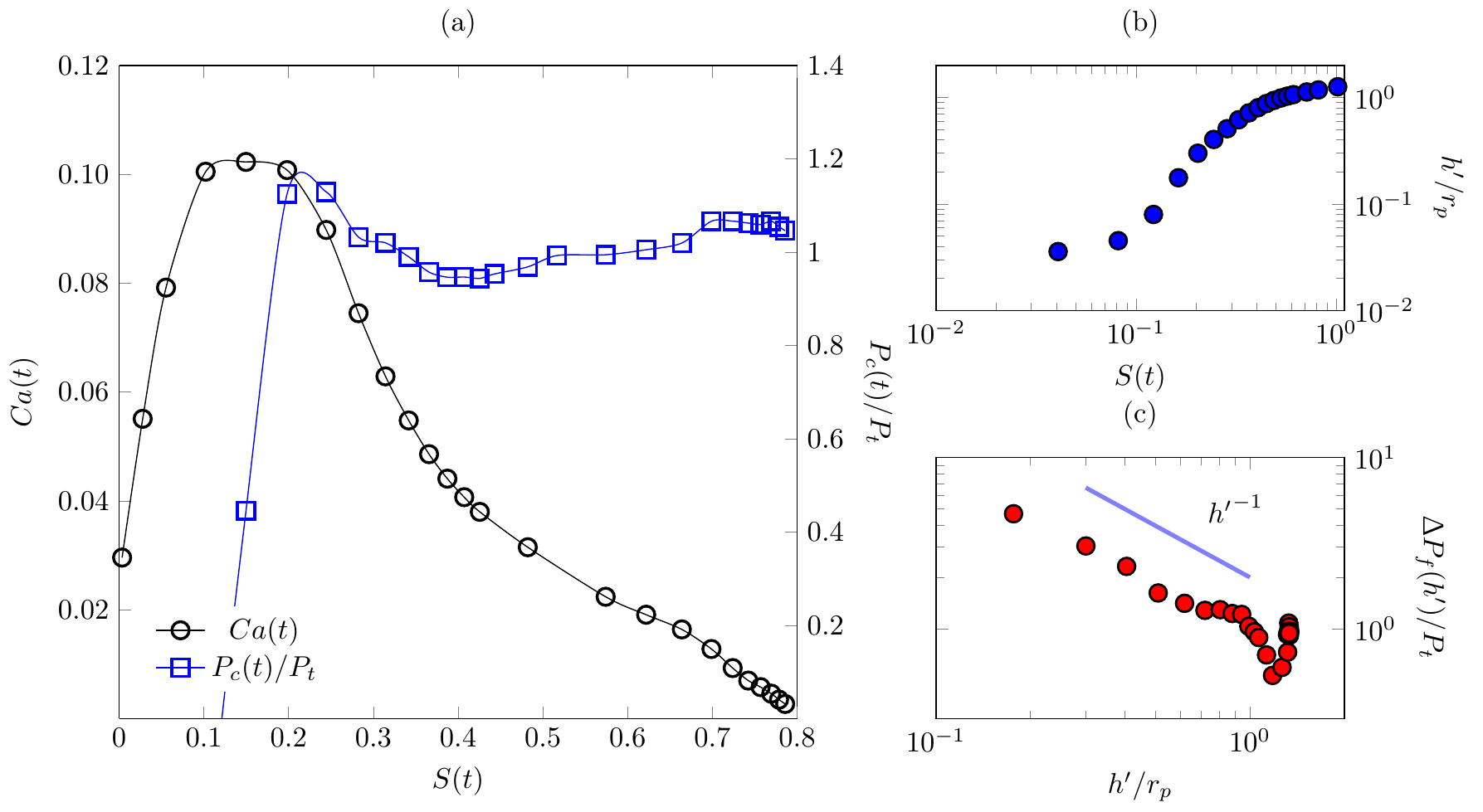}
        \caption{Right panel (a): capillary pressure and capillary number plotted as functions of water saturation. In the inertia-dominated region, $S<0.2$, inertial effects dominate, the liquid accelerates and the capillary pressure sharply increases, i.e. $\mathrm{d} Ca /{\mathrm{d}S}>0$ and $\mathrm{d} P_c /{\mathrm{d}S}>0$. For $S > 0.2$, the flow dynamics is characterized by both capillary and viscous resistance forces. In this regime we observe, $\mathrm{d} Ca/\mathrm{d}S<0$ and $\mathrm{d} P_c/\mathrm{d}S\sim0$. Left panels:  (b) front width $h'$ as function of time, and (c) the non-dimensional pressure drop along the front width  $\Delta P_f/P_t$  as function of the front width $h'$}
    \label{fig:pc_s}
\end{figure}

Fig.~\ref{fig:pc_s} presents the variation of capillary pressure and the values of the capillary number $Ca(t)=\mathrm{d} h/ \mathrm{d} t \ \mu_{w}/{\sigma}$ during the liquid infiltration. The capillary pressure is non-dimensionalised with the estimated capillary pressure based on the average pore size $P_t$. The computed values are plotted as a function of the saturation of the medium.
In the early stage, the capillary pressure increases as the liquid enters the GDL, i.e. $\mathrm{d}P_{c}/\mathrm{d}S>0$. We can imagine this increment as a result of the initial inertial stage, where the strong dynamics induces incrementally larger curvatures of the interface. This argument is further corroborated by the increasing value of the capillary number.

After the initial growth and after having reached a maximum, the capillary pressure slightly decreases. The observed maximum at $P_c \sim 1.2 P_t$ suggests the presence of pore throats smaller than the average pore size, which induce high interface curvatures and capillary pressures, when the liquid front has advanced, on average, roughly the 20\% of the medium length. To the observed decrease of the capillary pressure at $S(t)>0.2$ corresponds a decrease of the liquid infiltration rate, quantified by the capillary number $Ca(t)$, and, in turn, the beginning of the regime where viscous forces incrementally contribute to the flow resistance.
The regime transition occurs at a time comparable with the non dimensional viscous time $t_v^*\sim 0.64$. However,the viscous time characterising such a transition can significantly vary depending on various conditions defining the two-phase system, even in simple geometries, and thus the formulation of $t_v$ chosen here offers us only an approximated indication of the expected flow behaviour (see e.g.~\citet{stange2003} for an interesting discussion on the topic). We indeed have to consider the additional effect that the microstructure could induce on the partitioning between inertial, viscous and capillary forces.
We then argue that the transition between flow regimes is likely to be also induced by a significant geometrical constriction, i.e. by the presence of a significant amount of small pore throats. Given the observed variation of pore radius values in the samples (see Fig.~\ref{fig:psd}), we expect both small and larger pore throats delineating the morphology of such a constriction.
The liquid phase breaks through the largest pore (with the smallest capillary resistance), the interface curvature is reduced and in turn the capillary pressure. The infiltration velocity also decreases as a consequence of the increased viscous resistance.
By looking at the observed macroscopic behaviour of the liquid infiltration shown in Fig.~\ref{fig:pipe} we immediately recognize that to $S(t)\sim 0.2$ approximately corresponds the transition from the ballistic dynamics to the later regime $S(t)\propto {t^*}^{1/2}$. In the next section we will further evaluate the impact of the GDL morphology on the dynamic transition between two-phase flow regimes.

In the late stage, the capillary pressure approaches a constant value of $P_{c} \sim  \ P_t$ while the capillary number tends to zero. The latter observation does not imply that the liquid flow is null, but rather that the saturation does not change in time and the liquid follows preferential paths, so that $Ca(t)\propto \mathrm{d} h/\mathrm{d}t\rightarrow 0$. The initial trend suggests that the capillary pressure experiences a dynamic behaviour, which results proportional to the average velocity of infiltration and $\mathrm{d}P_c/\mathrm{d}S \propto \mathrm{d}Ca/\mathrm{d}S$, as already observed in literature (see e.g.~\citep{weitz1987}). It is important to stress that for $S(t) > 0.2$, both the viscous and capillary forces play an important role in determining the liquid motion, so that this regime can be referred to as a capillary-viscous one. 

In order to further characterise the two-phase interface displacement, we estimate the front width (or roughness) as the root mean square of the front fluctuations as $h'(t)=(\sum_A (h^*(y,z,t)-h(t))^2/(A\phi))^{1/2}$ (with $h^*$ the local front position along $x$). We observe that in the viscous-capillary regime, the front width increases in time, as can be observed in the Fig.~\ref{fig:pc_s} (b). We also observe that in such a flow region the pressure drop along the front width $\Delta P_f (h')$ (between the base $h-h'$ and tip $h+h'$) decreases almost linearly with $h'$, in Fig.~\ref{fig:pc_s} (c). These two observations suggest that, after the initial invasion event, the infiltration occurs in an unstable manner. A decrease of the pressure difference between base and tip of the front indeed suggests that, as the front becomes rougher, the probability of invasion at the tip increases with respect to that at the base, promoting unstable front displacement. In another equivalent perspective, the viscous resistance (pressure drop) along the front width decreases during kinetic roughening. A similar destabilising mechanism has been discussed for instance by~\cite{xu1998}. In the next section we will observe that it is the microstructure variation along the through-thickness direction that is responsible for such a peculiar front dynamics.

\subsection{Pore-scale water spatial distribution within the GDL\label{sec:pore}}

\begin{figure}
    \centering
    \includegraphics[width=0.8\linewidth]{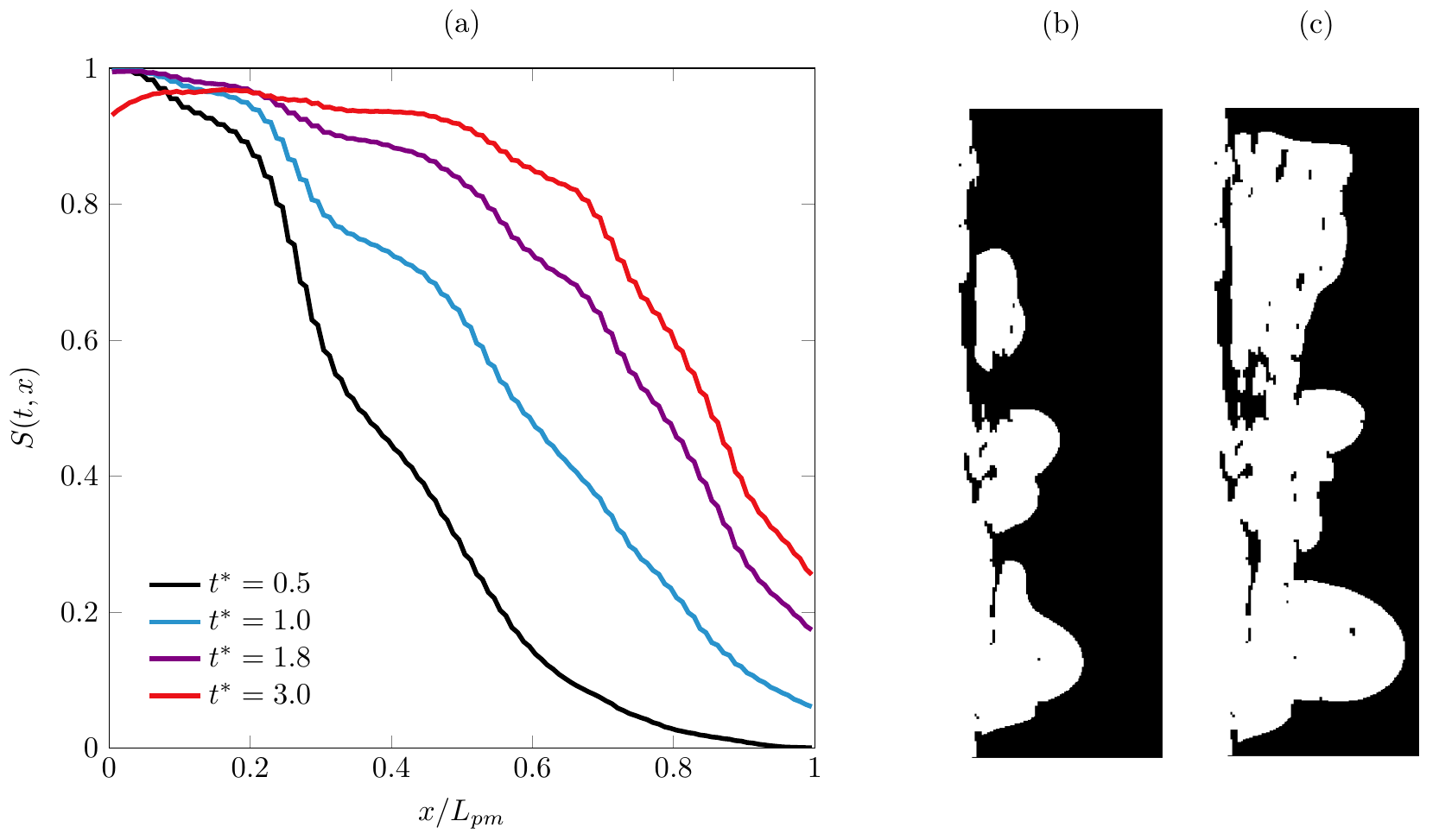}
    \caption{(a) Cross-sectional saturation along the domain thickness at various times $t^{*}$ after the breakthrough. The value of saturation diminishes along the flow direction $x$ indicating that less and less portions of the cross sections are occupied by the liquid phase. On the right panels, a snapshot is presented of liquid-gas distribution in a $(x-y)$ plane at dimensionless times $t^*=0.5$ (b) and $t^*=1.8$ (c)}
    \label{fig:sat}
\end{figure}

Pore-scale numerical simulations provide much deeper insights than do the models based on macroscopic quantities. As it can be observed from Fig.~\ref{fig:invasion}, the spatial distribution of liquid in GDLs results heterogeneous and, as discussed in the previous section, the front exhibits kinetic roughening. The heterogeneous redistribution of water inside the porous medium can be observed along the primary direction of the flow, as shown in Fig.~\ref{fig:sat}. The liquid phase does not penetrate uniformly, but rather the liquid front presents a smooth shape, as indicated by the decrease of the saturation along the thickness of the medium.
This decrement indicates that, after the breakthrough time $t^*\ge 0.5$, the liquid phase exhibits a compact uniform front at the GDL entrance, whereas farther along the GDL, it is redistributed along different preferential flow paths (see also the right panels of Fig.~\ref{fig:sat}). These preferential finger-shaped paths are fluid structures typically observed during two-phase fluids invasion in porous media. For instance, viscous fingering refers to the situation when a less dense fluid displaces a denser one and, for sufficiently high invasion velocities, viscous forces destabilise the invading front and sustain the formation of fingers at the interface~\citep{meheust2002}. On the contrary, when a denser fluid invades a medium initially filled with a less dense one, the viscous forces tend to stabilise the front~\citep{aker2000}. In fuel cells the GDL is usually initially filled with a gaseous fluid (hydrogen or oxygen). As the cell starts operating, liquid water (more dense) is produced and invades the medium. In such a situation, with $M=\mu/\mu_g\gg1$, viscous forces contribute to stabilise the front while the distribution of pore-scale capillary thresholds determines the onset of fingers, a phenomenon known as capillary fingering. The occurrence of capillary fingering is known to depend also on other physical parameters, such as the capillary number and the hydrophobicity of the material, the latter determining the shape of the liquid-gas curvature at the pore throats~\citep{lenormand1988,cottin2010}.

For instance, the transition from uniform infiltration to capillary fingering has been observed in media with uniform randomly distributed pore sizes, when changing the liquid-gas-solid contact angle, from hydrophilic to hydrophobic~\citep{jung2016}, or when controlling the competition between pore size gradients and medium disorder~\citep{lu2019}.
Here, we observe a similar transition, but strongly governed by the porous microstructure. The high value of the capillary pressure recorded (see Fig.~\ref{fig:pc_s}) can be interpreted as the result of an ``effective'' contact angle, which, rather than by the chemical properties of the solid surface (have in mind that in our simulations the equilibrium contact angle is $\theta_{eq}=90^\circ$), is induced by the combination of strong dynamics of the capillary pressure and the presence of significant local constrictions of the flow path, which leads to a high curvature of the two-phase interface. Subsequently, the interface experiences a mechanical instability and a successive sudden invasion event occurs across certain pore throats. Such an effective contact angle appears then to induce noncooperative
burst instabilities that dominate the interface advance to form ramified fluid structures, as observed for porous media with contact angles $\gtrsim 120^\circ$~\citep{jung2016}. In other words, the invading fluid can be considered as ``non-wetting'' as we observe the occurrence of capillary fingering even with non-hydrophobic surfaces. The mechanism of microstructure-induced capillary thresholds counteracts the viscous forces to destabilise the front as also confirmed by the scaling of the pressure drop at the front width observed in Fig.~\ref{fig:pc_s}.
In Section~\ref{sec:model} we will quantify such an effect through the formulation of an effective model. A similar physical approach, which makes use of an effective contact angle, has been used for example for the understanding and characterisation of two-phase flow dynamics in soils, see e.g. ~\citet{wessel1988}.

\begin{figure}
    \centering
    \includegraphics[width=0.7\linewidth]{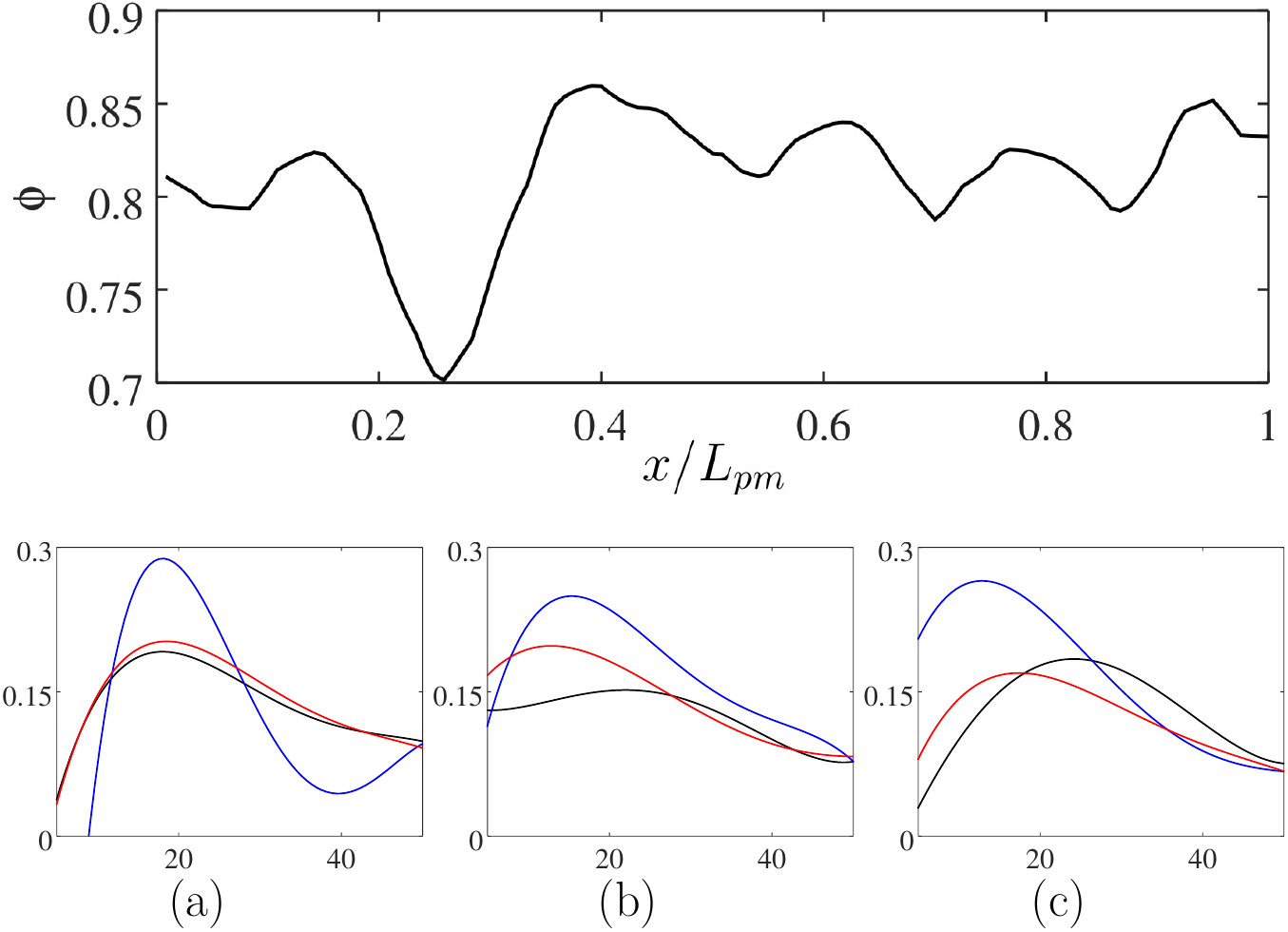}
    \caption{Porosity, $\phi$, as a function of GDL normalized thickness $x/L$ (top) and the probability distribution functions (PDFs) of pore radii at three cross sections (bottom) for the three samples. In the bottom figure, the vertical and horizontal axes are the relative frequency and pore radius, respectively. The PDFs are presented at $x/L_{pm} \sim 0.1$ (black lines), $x/L_{pm} \sim 0.25$ (blue lines) and $x/L_{pm}\sim 0.8$ (red lines)}
    \label{fig:por}
\end{figure}

In Fig.~\ref{fig:por} we report the variation of planar porosity along the domain. The variation of the porosity suggests the presence of pore constrictions followed by pore enlargements. Such a microstructural traits is likely to induce a positive permeability gradient along the flow direction, thus to destabilise the front after it invades the first pore constriction at $S\sim2$, and to induce the capillary fingering observed in Fig.~\ref{fig:sat} and the scaling depicted in Fig.~\ref{fig:pc_s}~\citep{xu1998}. This mechanism can also be intuitively understood by considering that the viscous pressure drop is inversely proportional to permeability (Darcy's law) and that thus a sufficiently high increase of the permeability along the flow direction can cause a decrease in the viscous drop along the front. It also confirms the important role that pore-scale statistical variations along the medium thickness can have in determining a fluid displacement, given the lack of separation of length scales characterising the pore and minor medium dimensions. Notably, we observe a minimum value of the porosity at $x/L_{pm}\sim0.25$, an observation that sustains our hypothesis that a significant change in the two-phase flow behaviour coincides with the occurrence of a change in the pores morphology (see also Fig.~\ref{fig:pc_s}).
In particular, the pore radius PDFs computed at different cross-sections depicted in the bottom of Fig.~\ref{fig:por} show the more frequent occurrence of small pore throats at $x/L_{pm}\sim0.2$. Such small pores of radius $\sim 15 \ \mu m$  are significantly smaller than the average pore radius $r_p=40 \ \mu m$.  We suggest that the presence of pores with a small radius in a certain region of the GDL can be due to an inhomogeneous distribution of the binder, which is used to bound together the polymer matrix during GDL fabrication. A bimodal distribution of pore sizes has been for instance observed  in reconstructed GDLs and associated with substantial differences in the binder amount in different regions~\citep{zenyuk2016}, a difference that here we observe to be significant in the through-plane direction.

 The transition occurs thus both in space and time: liquid invasion occurs rapidly and uniformly for short times, until the front reaches $x/L_{pm}\sim 0.2$ (just before the observed constriction at  $x/L_{pm}\sim0.25$). Later in time and farther in space, the liquid is divided into clusters; infiltration occurs firstly on the large pores and then along various flow paths or fingers in an unstable manner.

\subsection{The structure of the liquid phase emerging from the GDL\label{sec:fing}}

\begin{figure}
\centering
\includegraphics[width=0.7\linewidth]{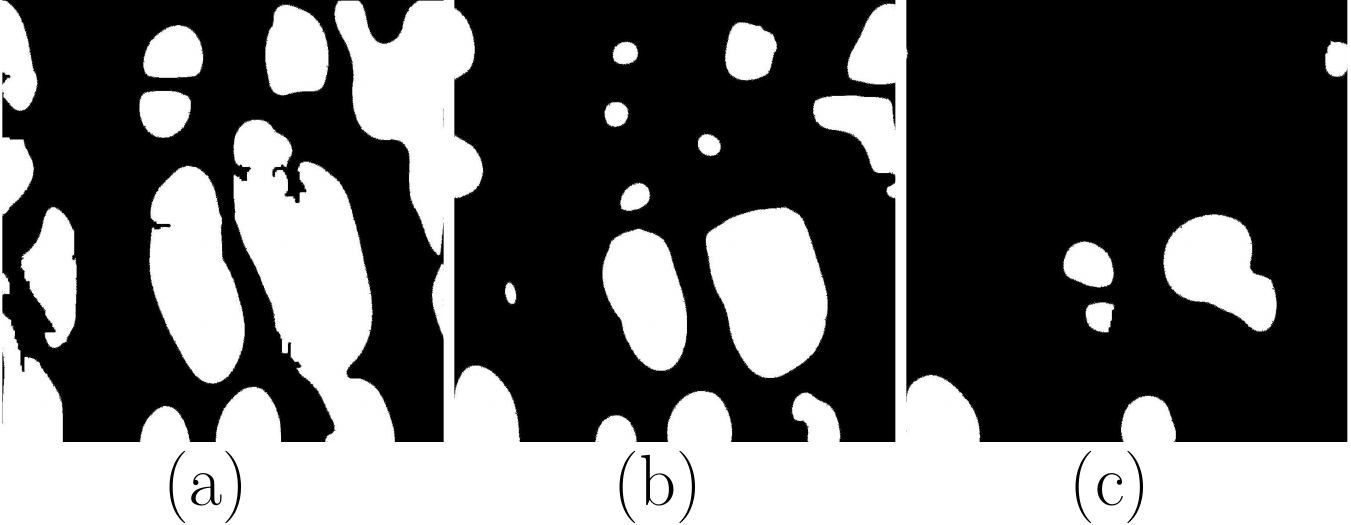}
\caption{Liquid distribution at the breakthrough time at three slices along the streamwise direction for sample 3; (a) $x/L_{pm}=0.4$, (b) $x/L_{pm}=0.6$ and (c) $x/L_{pm}=0.8$. The white domains represent liquid clusters (fingers), while the black ones are solids or gas. The simulations illustrate that the invading fingers of the liquid water become thinner as the liquid advances in the porous medium.} \label{fig:finger}
\end{figure}

The breakthrough time defines the moment when the first portion of liquid reaches the outlet. It is then interesting to observe the spatial configuration of the liquid phase at this instant, since it provides information about how the porous structure has affected the mass transport from the inlet to the outlet of the medium. We should also emphasise that the breakthrough time ($t^*\sim 0.5$) occurs shortly after the transition between the ballistic and viscous-capillary regimes ($t^*\sim 0.3$), an insight that suggests a fast dynamics of pore invasion after the liquid front overcomes the capillary barrier induced by the pore constriction at $x/L_{pm}\sim 0.2$.

\begin{figure}
    \centering
    \includegraphics[width=0.9\linewidth]{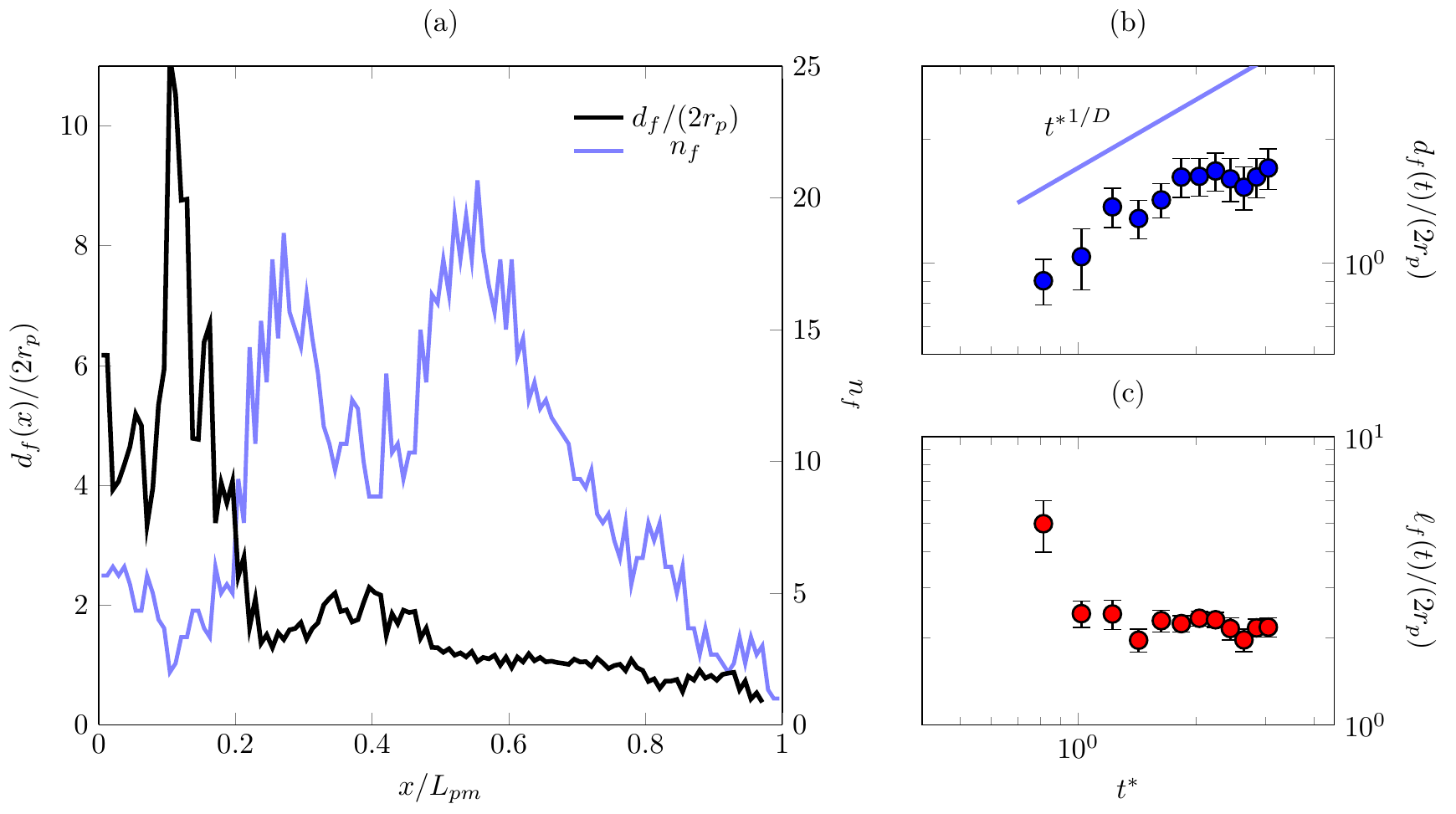}
    \caption{Left panel (a): cross-sectional average fingering diameter $d_f$ and number of fingers $n_f$ in the through-thickness direction at breakthrough time ( $t^{*}=0.5$).
The average diameter and the number of liquid clusters or fingers $n_f$ decrease as the liquid evolves along the GDL thickness. Note that the initial part of the GDL ($x/L_{pm} <0.2$) is almost uniformly filled with water. Right panels: average fingering diameters $d_f$ (b) and  and average finger distance $\ell_f$ (c) on the outlet surface of the GDL $(x/L_{pm}=1)$ as function of time after breakthrough $t^*\ge0.5$. The finger dimension shows a time-dependent increase, suggesting $d_f\propto {t^*}^{1/D}$}
    \label{fig:finD}
\end{figure}

In Fig.~\ref{fig:finger} the spatial configuration of the liquid fingers at various cross sections after the constriction is shown for one of the REVs at the breakthrough time $t^*\sim 0.5$. After $x/L_{pm}\sim0.2$, the average dimensionless diameter $d_f/(2r_p)$ and number  $n_f$ of the fingers gradually decrease along the cross-sections of the medium. This trend is quantified in Fig.~\ref{fig:finD}, where the latter quantities are plotted along the medium thickness. We firstly note that the initial part of the GDL ($x/L_{pm} <0.2$) is almost uniformly filled with water, since $d_f^2 \pi /4  \ n_f \sim A\varepsilon$. In this region, rather than fingers, we observe clusters of liquid separated by fibers, with the medium almost fully filled, as also shown in Fig.~\ref{fig:sat}, where $S(x/L_{pm} <0.2)\sim 1$.

Closer to the GDL outlet, the diameter of the fingers decreases from an average size of $d_f\sim 160 \ \mu m = 4r_p$ at $x/L_{pm} \sim 0.5$, to $d_f\sim 80 \ \mu m = 2r_p$ at $x/L_{pm} \sim 1$. In Fig.~\ref{fig:finD}, a decrease of the number of fingers along the GDL thickness is also observed.
The quantification of the liquid structures thus offers us another important insight: at the breakthrough time, the distribution of liquid invasion gradually changes in space along the medium, from a collaborative pore filling behaviour, with clusters of the size of few pores $d_f(x/L_{pm} \sim 0.5)\sim 4 r_p$, to a preferential flow along a single pore, when $d_f(x/L_{pm} =1)\sim 2r_p$, as more and more pore throats contribute to impede liquid transport along the direction of the flow.  Interestingly, this finding motivates a possible new design of GDLs, as characterized by an increasing size of the pore diameters, as a recent study also suggested~\citep{balakrishnan}. Such a design should promote the water transport in the GDL by decreasing the capillary pressure thresholds, induced by the pore throats, along the flow direction. However, such an innovative design should also be carefully tailored to take into account the effects of the spatial distribution of pore sizes (i.e. the amount disorder of the medium), a parameter which is found to be important for the control of capillary fingering phenomena~\citep{lu2019}.

\begin{figure}
    \centering
    \includegraphics[width=0.7\linewidth]{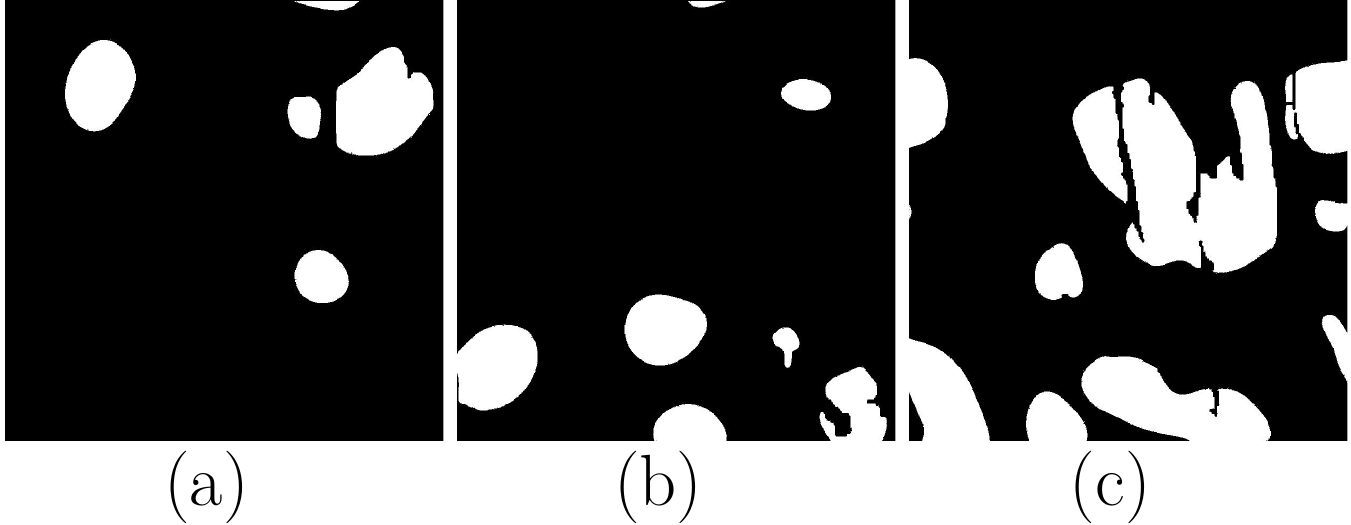}
    \caption{Liquid distribution at the outlet surface for the three samples at $t^{*} \sim 2$}
    \label{fig:fing_outlet}
\end{figure}

The observed spatial redistribution of the liquid phase from clusters to single invaded pores at the breakthrough time suggests that water primarily emerges from individual pores distributed on the GDL outlet surface.
To better understand the characteristics of the structures characterising the liquid transport along such preferential paths, from the inlet to the outlet of the GDL, we quantify the shape and dynamics of the liquid fingers at the outlet cross-section, after the breakthrough time, for $t^*\ge 0.5$. The computed values of $d_f$ are reported in Fig.~\ref{fig:finD}, where the average distance between emerging fingers $\ell_f$ is also computed, while in Fig.~\ref{fig:fing_outlet} the emerging fingers at the outlet cross-section are displayed for a visual comparison. As previously suggested, the liquid phase emerges in the form of fingers with the size comparable with the average pore sample diameter $\sim 2r_p$. In the times following the breakthrough, the average size of the fingers slightly increases, as well as their number (not shown here), as more liquid reaches the outlet of the medium along a gradually increasing number of pores. 

The time-dependent increase of finger dimension on the outlet surface suggests a power-law scaling $d_f\propto{t^*}^{1/D}$, being $D$ the fractal dimension of the emerging cluster. This observation is consistent with classic percolation-invasion studies that predict the more probable invasion site occurring at a distance $\propto {t^*}^{1/D}$ from the previous event, with $D\sim 1.82$~\citep{furuberg1988,wilkinson1983}. A least-square fit would give us $D\sim 2.3$, even though we note that such a scaling is limited to just half a decade. 
On the contrary, we find that the occurence of emerging fingers on the GDL surface appears to be uncorrelated, since their average computed distance fluctuates around a constant value $\ell_f\sim 2r_p$, as reported in the bottom-right panel of Fig.~\ref{fig:fing_outlet}.

\subsection{Modelling water transport in the GDL~\label{sec:model}}

We have observed three regimes that characterise the mass transport within the GDL. Among them, the ballistic regime occurs at short times and only in the entrance region of the GDL. In this section we aim at modelling the water transport from a macroscopic perspective and we therefore delineate a model based on the long time regimes that dominate the great part of the two-phase flow dynamics in GDLs.

During the viscous-capillary stage, the liquid flows through an increasing number of preferential pores and it experiences an increasing viscous resistance induced by the augmented liquid-solid interface. As observed in Fig.~\ref{fig:pipe}, the liquid transport globally follows a power law $S(t)\propto {t^*}^{1/2}$ until a significant amount of water has reached the outlet of the GDL and $\mathrm{d} S/\mathrm{d} t \sim 0$.
We can express the transport of water inside the GDL until this time ($t^*\sim 2$) by means of a model based on Darcy's formulation. The latter assumes a linear relation between the pressure drop and the flow rate, as expressed in equation~\ref{eq:darcy}, which is quantified by the permeability of the medium. Darcy's equation describes the flow in terms of average values, which are obtained  at length scales much larger than the structural heterogeneities of the porous materials. Therefore, a model based on Darcy's equation must inevitably rely on effective quantities that convey information from micro- to macro- scales.
To build up a Darcy-based model, we then extract the effective quantities that govern the dynamics of liquid infiltration in the capillary-viscous regime, where we observe a Darcian behaviour of the fluid, i.e. $S(t)\propto {t^*}^{1/2}$. In this regime, the flow dynamics is induced by the pressure drop $-\nabla_xP L_{pm}$ acting on the average penetration length $h$; the capillary pressure induced by the porous microscructure adds an additional resistance to the flow, which cannot be neglected, as discussed in Section~\ref{sec:liq}. We can then define an effective permeability $K_{\mathit{eff}}$ that embeds the effect of the microstructure and rewrite equation~\ref{eq:darcy} as:
\begin{equation}
\label{eq:capillary}
\frac{\mathrm{d} h}{\mathrm{d} t} = -\frac{K_{\mathit{eff}}\ \nabla_x P\ L_{pm}}{\mu_w\ h} \ ,
\end{equation}
where $K_{\mathit{eff}} = 1.2 \ 10^{-10}\  m^{2}$ is the effective permeability extracted by fitting the values reported in Fig.~\ref{fig:pipe}. The same equation can be rewritten in a dimensionless form as:
\begin{equation}
\label{eq:capillary_non}
S(t)=\big ( K_{\mathit{eff}}^{*} \  t^{*} \big )^{\frac{1}{2}} \ ,
\end{equation}
with  $K_{\mathit{eff}}^* = K_{\mathit{eff}}/(r_p^2/4)= 0.30$, which points out that, from a macroscopic perspective, the liquid infiltration occurs similarly to the infiltration in a pipe with an effective radius $ r_{p,\mathit{eff}} \sim 0.55\ r_{p} \sim 22 \ \mu m$. The value of the effective radius, lower than half of the observed average finger diameter, indicates the adverse effects of the capillary pressure on the flow dynamics, which is then characterised by an increased resistance to the flow. Recalling the picture of an effective contact angle induced by the microstructure discussed in Section~\ref{sec:pore}, we can estimate this angle from the value of the calculated effective radius. The liquid-gas curvature induced by the microstructure provides an additional resistance to the flow in the form of an effective capillary pressure $P_{c,\mathit{eff}} = -\nabla_x P L_{pm} (1-r_{p,\mathit{eff}}^2/r_p^2)$. This value ($P_{c,\mathit{eff}} \sim 0.7\ P_t$) is very close to the observed value of capillary pressure at long times (see e.g. Fig.~\ref{fig:pc_s}). From the effective capillary pressure, the effective contact angle is then calculated as:
\begin{equation}
\theta_{\mathit{eff}} = \arccos \bigg (-P_{c,\mathit{eff}} \frac{r_p}{2\gamma} \bigg ) \approx 134^\circ \ ,  \label{eq:contact}
\end{equation}
a value that we notice places the two-phase flow dynamics  in the regime of noncooperative
burst instabilities for two-phase flows in porous media~\citep{jung2016}.

It is interesting to observe that by using the calculated effective contact angle or, equivalently, the effective pore radius, we determine the characteristic invasion velocity as $U_{c,\mathit{eff}}= r_{p,\mathit{eff}}^2 / r_p^2 \sim 0.3\ U_c$ and the characteristic time for liquid to reach the GDL outlet as $t_{c,\mathit{eff}}\sim 3\ t_c$.
Indeed, at around this time, for $t^*> 2$, the mechanism changes by which water is transported, since enough preferential flow paths connecting the inlet and outlet of the medium are formed. In this later stage, the liquid is transported preferentially along the liquid fingers and the total saturation of the medium approaches a constant value $S\sim 0.75$. We can estimate the volumetric flow rate at the outlet in this later stage by means of the product between the average liquid velocity  $u_f = -\pi d_f^2\ \nabla_x P / (16 \mu_w)$ and the liquid area at the outlet $A_f=n_f d_f^2 \pi /4$. The calculation leads to a volumetric flow rate $q_{out}=u_f\ A_f \sim 5 \ cm^3/s$,  and a capillary number $Ca_{out}(t) \sim 0.5$,  at the time $t^*=2.5$. These are useful quantities for the prediction of water emerging from the GDL and transported within the fuel cells gas channels.

In a macroscopic effective model that predicts the two-phase mass transport in fuel cells, Eq.~\ref{eq:capillary} can describe the motion of the fluid in the bulk of the GDL, via an effective contact angle given in Eq~\ref{eq:contact}. Furthermore, the calculated flow rate and the capillary number at the outlet can serve as microscopic input parameters for the boundary conditions of transport at the GDL-gas channels interface. As a final note we point out that the value of $Ca_{out}(t)$, which is close to unity, underlines the fast dynamics that water experiences when the liquid reaches the GDL outlet and, in turn, the prominent role that inertia plays in the formation of liquid droplets emerging from the GDL surface.

\section{The transport of large respiratory droplets within fibrous porous microstructures~\label{sec:covid}}

Because of the urgent need for sharing research results and information in a global effort to deal with the current COVID-19 pandemic, we finally want to highlight the significance of the present results for the understanding and development of effective personal protective equipment (PPE). Among them, face masks play an important role for source control (used by infected persons) or prevention of COVID-19 (used by healthy persons), whether they are surgical masks used in health care environments or nonmedical masks used in the community
setting~\citep{world2020}. It is crucial to optimise the face masks effectiveness in filtering pathogen-bearing respiratory droplets, without significantly compromising their comfort and accessibility to the public (cost). A possible solution lies in the identification of the optimal microstructural design of the masks' filter layers, which in turn facilitates both the decision-making process to issue recommendations with regards to wearing certain types of masks, and the production of new low-cost effective materials. The fibrous porous structure of the gas diffusion layer investigated here strongly resembles many common fabrics materials that can be used as PPE, being composed of layers of interconnecting fibres.

For the sake of clarity, we here define ``large respiratory droplets'' as the large pathogen-bearing droplets characterised by diameters $d_b$ significantly larger than the mean pore size of the filter layer, that is, with $d_b \gg 2r_p$. In such a way, they can be well represented by a liquid front impacting a surface area of the size of the here investigated  samples, $A \approx (20 r_p)^2$. For instance, for face masks with a mean pore size of $\approx$ 50 $\mu$m, the size of the large respiratory droplets would be  $\approx$ 500 $\mu$m, a value that characterises a small yet important portion of cough-generated droplets, since this portion is more likely to contain pathogenic organisms~\citep{duguid1946}.
The present results unveil three very important insights about the transport of large pathogen-bearing through filter layers of face masks.

(i) The fibrous microstructure, even if not hydrophobic, induces an effective resistance to the their transport. Such an effective resistance, which we quantified via an effective contact angle in Eq.~\ref{eq:contact}, is particularly important when the expiratory event induces a high value of the pressure difference between the two sides of the mask, and the resulting pressure drop is comparable with the characteristic capillary pressure of the microstructure $P_t=2\sigma/r_p$. Such a scenario is for instance encountered when a mask-wearing subject coughs and the oesophageal pressure exhibits very high values, in the order of tens of kilopascal~\citep{lavietes1998}.

(ii) Since, as we have observed, pore throats incrementally contribute to impede liquid transport, the design of a layered porous structure with a decreasing size of the pore diameters along the flow direction can augment the filtration mechanism of large respiratory droplets, without necessarily increasing the thickness of the filter layer. Since the filtration mechanism in protective masks can be important both in inward and outward directions (e.g. during inhaling and exhaling), such a gradient in pore sizes could be designed as function of thickness relative to the middle plane of the porous material. A similar design should however also take into account the microstructure-induced pressure drop and thus the provided comfort (breathability). The existence of an effective capillary pressure opposing the liquid transport in neutrally wetted porous media and the expected positive effect of microstructures with variable pore sizes can possibly explain why layered designs of commonly available fabric materials, devoid of protective hydrophobic layers, have been found to offer protection about equivalent to the filtration offered by 5-layer N95 masks~\citep{lustig2020}.

(iii) Finally, we also observe the emergence of the liquid phase from the outer surface of the porous medium, in the form of fingers with diameter sizes comparable with the mean pore size of the material. These fingers are rather homogeneously spatially distributed and their characteristic velocity is quantified via $Ca_{out}\sim 0.5$. Such an information is important for both the possible targeted design of masks with acceptable limits in the size of transmitted droplets, and for the delineation of health risk assessments related to the expected size and velocity of pathogen-bearing droplets expelled from face masks.

\section{Conclusions\label{sec:end}}

We have presented lattice Boltzmann simulations of two-phase flows performed in a CT reconstructed fuel cell gas diffusion layer (GDL). We have been able to identify the pore-scale characteristics of the two-phase flow dynamics and spatial distribution within the porous medium, with a micrometre accuracy. We have sampled three different regions of the reconstructed material to allow a statistically robust analysis.

We have identified three regimes that characterise the two-phase flow dynamics in GDLs. At short times, for $t^*\le 0.3$, the liquid phase rapidly invades the medium rather uniformly with saturation following a ballistic behaviour $S(t)\propto {t^*}^2$. When the liquid front reaches the first significant flow constriction, represented by an increase in the number of small pore throats, it exhibits a different behaviour. We have observed that to such a change in the porous morphology (presumably related to a local increase of the binder in the gas diffusion layer) corresponds a maximum value of the computed capillary pressure. The capillary pressure is found to act adversely to the flow, being induced by a convex shape of the liquid-gas interface when approaching the pore throats. Such an effect is observed in a three-phase (liquid-gas-solid) system with a neutral contact angle and it is therefore purely induced by the interplay between flow dynamics and microstructure. From this moment, at intermediate times $0.3 < t^* \le 2.5$, the liquid phase breaks through certain pores and forms preferential flow paths of the size of one-two pores. We have observed the liquid phase to be transported mainly through the mechanism of capillary fingering during this stage, with an increasing number of preferential flow paths contributing to the global transport. During this stage, the invasion dynamics exhibits an unstable displacement and a kinetic roughening of the front, which is sustained by the sudden enlargement of the pore throats. At the same time, we have observed a gradual decrease of the liquid velocity as a consequence of the gradually increasing liquid-solid surface that provides viscous resistance, so that the saturation follows a  power law $S(t)\propto {t^*}^{1/2}$. At long times, for $t^*> 2.5$, when a sufficient amount of liquid has reached the outlet, the two-phase fluid structure stabilises and the liquid flows exclusively following the previously formed preferential flow paths.

On the basis of these observations, we have proposed a model to mimic and predict the liquid transport along the gas diffusion layer. By making use of an effective contact angle that quantifies the effect of the measured capillary pressure, we have been able to build up a simple equation of motion, based on Darcy's equation, that well approximates the transport of water within the fuel cell material. We argue in the paper that such an effective contact angle does not only explain the noncooperative pore filling and unstable front displacement behaviours observed but also the strong intermittent dynamics of two-phase flows and pressure signals often measured in fuel cells during flooding conditions.

In the last part of the manuscript, we have highlighted the significance of these results for the optimal design of filter layers of protective masks. We have explained the mechanisms governing  the transport of large respiratory droplets, produced by intensive expiratory events, through fibrous porous microstructures. We have discussed how such forms of pathogens transmission can be efficaciously mitigated also through the use of nonmedical face masks made of common fabrics materials, provided that the proper characteristics of the different layers that compose them are selected.

\section*{Declarations}

\subsection*{Funding}
The research project to which this work belongs has received funding from the European Union's Horizon 2020 research and innovation programme under the Marie Sklodowska-Curie grant agreement No 790744 and from the Chalmers' Areas of Advance Transport.
The computations were enabled by resources provided by the Swedish National Infrastructure for Computing (SNIC).

%
 \subsection*{Conflicts of interest}
The authors declare that they have no conflict of interest.

\subsection*{Availability of data and material} 

An excerpt of data is available on the EUDAT B2SHARE platform at \url{https://b2share.eudat.eu/records/32e06d230539414d8afb146f595ca6ea} under a Creative Common Attribution-NonCommercial-ShareAlike 4.0 International License.

\subsection*{Code availability}

The code used for the numerical simulations is accessible at \url{https://gitlab.com/dariom/lbdm} under a Creative Common Attribution-NonCommercial-ShareAlike 4.0 International License.

\bibliographystyle{spbasic}      
\bibliography{manuscript}   

%
%

\end{document}